\documentclass[aip, jmp, a4paper, amsmath, amssymb,
    reprint, onecolumn, twoside, nofootinbib]{revtex4-2}

\usepackage[utf8]{inputenc}
\usepackage[T1]{fontenc}
\usepackage{mathptmx}
\usepackage{bm}

\usepackage{stmaryrd}    
\usepackage[cal=boondoxo,scr=pxtx]{mathalfa}
\usepackage{amsthm}
    \theoremstyle{plain}
        \newtheorem{theorem}{Theorem}
        \newtheorem{property}{Property}
        
        \newtheorem{lemma}[theorem]{Lemma}
        
    \theoremstyle{definition}
        \newtheorem{definition}{Definition}
        \newtheorem*{notation}{Notation}

    \theoremstyle{remark}
        \newtheorem*{remark}{Remark}

\usepackage[colorlinks,hyperindex,unicode]{hyperref}
\hypersetup{
   colorlinks,
   citecolor=blue,
   linkcolor=blue,
   urlcolor=blue,
}
\usepackage{orcidlink}

\newcommand{\llbra}{\llbracket}
\newcommand{\rrket}{\rrbracket}
\newcommand{\sgn}{\mathrm{sgn}}
\newcommand{\Lie}{\ensuremath{\mathscr{L}}}
\newcommand{\sss}[1]{\scriptscriptstyle{#1}}
\newcommand{\dr}{\partial}
\renewcommand{\S}{\ensuremath{\mathscr{S}}}
\newcommand{\Id}{\ensuremath{\mathrm{Id}}}

\newcommand{\Riemann}{\mathrm{Rm}}
\newcommand{\Ricci}{\mathrm{Ric}}
\newcommand{\R}{\mathcal{R}}
\newcommand{\csch}{\mathrm{csch}}
\newcommand{\sech}{\mathrm{sech}}

\renewcommand{\dot}[1]{\overset{\,_{\mbox{\Large .}}}{#1}}

\def\L{\Lie}
\def\LD{\Lie_{\sss\! D}}
\def\LF{\Lie_{\sss\! F}}
\def\iD{i_{\sss\,\! D}}
\def\iF{i_{\sss\,\! F}}
\def\iEf{i_{\sss\,\! E_f}}
\def\Fd{\mathbf{F}^2}

\def\setR{\mathbb{R}}
\def\setZ{\mathbb{Z}}
\def\calC{\mathcal{C}}
\def\calP{\mathcal{P}}
\def\calZ{\mathscr{Z}}
\def\calF{\mathscr{F}}

\def\R{\mathscr{R}}
\def\Id{\mathrm{{Id}}}
\def\Tr{\mathrm{Tr}}

\begin{document}
\title{FLRW embeddings in $\setR^{n+2}$, differential geometry and conformal photon propagator}

\author{E.~Huguet \orcidlink{0000-0002-0537-7750}}
    \affiliation{Universit\'e Paris Cité, APC-Astroparticule et Cosmologie (UMR-CNRS 7164), 
    Batiment Condorcet, 10 rue Alice Domon et L\'eonie Duquet, F-75205 Paris Cedex 13, France.}
    \email{huguet@apc.univ-paris7.fr}
\author{J.~Queva \orcidlink{0000-0001-8280-1925}}
    \affiliation{Universit\'e de Corse -- CNRS UMR 6134 SPE, Campus Grimaldi BP 52, 20250 Corte, France.}
    \email{queva@univ-corse.fr}
\author{J.~Renaud \orcidlink{0000-0002-3284-5003}}
    \affiliation{APC-Astroparticule et Cosmologie (UMR-CNRS 7164), 
    Batiment Condorcet, 10 rue Alice Domon et L\'eonie Duquet, F-75205 Paris Cedex 13, France.}
    \email{jrenaud@apc.in2p3.fr}

\begin{abstract}
    This paper introduces differential-geometric methods to study $n$-dimensional locally conformally  flat spaces as submanifolds in $\mathbb{R}^{n+2}$.
    We derive explicit formulas relating intrinsic and ambient differential-geometric objects, including curvature tensors, the codifferential, and Laplacian operators.
    We apply this approach to Friedmann-Lemaître-Robertson-Walker (FLRW) spaces using newfound embedding formulas, obtaining new and simplified expressions for the photon propagator in four dimensions.
\end{abstract}

\maketitle

\tableofcontents

\section{Introduction}

This article is twofold.
First, we devise differential-geometric methods to study $n$-dimensional locally conformally flat spaces by realizing them as submanifolds in $\setR^{n+2}$.
Second, using newfound embedding formulas for Friedmann-Lemaître-Robertson-Walker (FLRW) spaces, we readily apply this approach to such geometries.
In particular, we obtain new and simplified expressions of the photon propagator in these spaces in four dimensions. 
These expressions are made possible thanks to the clarity gained from this overlooking viewpoint in ambient space.

The conformal group of a $n$-dimensional conformally flat space is $SO(2,n)$, and its action is most naturally realized in $\setR^{n+2}$.
This makes this ambient space a well-suited background for studying conformally invariant fields.
To this end, Dirac\cite{Dirac:1936fq} introduced the manifestly covariant six-cone formalism for conformally invariant fields in Minkowski space, a framework that was later extended by Mack and Salam.\cite{Mack:1969rr}
A key ingredient, however, is an explicit embedding of the physical spacetime into the null cone of $\setR^{n+2}$.
The absence of explicit embedding formulas for generic conformally flat spacetimes accounts for why this approach has mostly been confined to flat and (anti--)de Sitter spacetimes [(A)dS],\cite{Binegar1982Mink,Binegar1982AdS,Bayen:1984dt,Huguet:2006fe,Huguet:2008js,Faci2009,queva:tel-00503186} 
in which it has been proven successful in the quantization of conformally invariant fields.
So we are faced with two problems if we follow this perspective for a broader class of spacetimes.
On the one hand, one needs to know, in a general manner, the relationship between ambient and intrinsic quantities.
On the other hand, to get tangible results, one needs explicit embedding formulas, preferably, as simple as can be.

Regarding the relationship between intrinsic and ambient quantities we note that it is fundamentally a matter of differential geometry.
For instance, one has to relate operators such as the Laplace--de Rham differential operator, viewed as the wave operator for the physical field, to its ambient realization.
There is then an interplay to analyze between restriction to and extension out of the embedded submanifold of differential operators.
Such a program has already been carried out successfully for scalar fields and one-forms,\cite{Huguet:2016szt,Zapata:2017gqg,Huguet:2022rxi,Huguet:2024dkw}
in particular to (A)dS spaces.
The current work encompasses these previous results, while vastly extending and simplifying them.
In our framework, the spacetime $X_f$ is obtained as the intersection of the null cone $\calC$ of $\setR^{n+2}$ with an hypersurface $\calP_f$, defined by a homogeneous function $f$ of degree one in ambient space.
We show, with explicit formulas, that every differential-geometric object needed such as the curvature tensors, the Laplace--de Rham $\square_f$, and related entities, are derived solely from this defining function $f$.

Regarding the need for explicit embedding formulas we first note that the question of embeddings of spacetimes is a question in itself.
These embeddings could serve a variety of goals.
For instance, a well-known embedding is that of (A)dS spaces, which can be viewed as hyperboloids in flat space,\cite{Birrell:1982ix,Weinberg:1972kfs} an embedding known to de Sitter himself.\cite{deSitter_1917,deSitter_1918}
This embedding can serve an illustrative role, helping to get a better understanding of the structure of this spacetime.
In addition, since the $SO(1,4)$ [or $SO(2,3)$] invariance of (anti--)de Sitter space is manifest in this ambient space this sets a proper framework, in $\setR^5$, to study fields in (A)dS spaces.\cite{Takook:2014paa,Pethybridge:2021rwf}
Fronsdal's\cite{Fronsdal:1959zza} embedding of the Schwarzchild metric is another well-known example allowing for a better understanding of the spacetime.
Higher-dimensional physics, such as Kaluza-Klein theories, is a major source of inspiration for investigating embeddings of (physical) spacetimes;
for a thorough compilation of references along these lines up to the year 2000, see Ref.~\onlinecite{Pavsic:2000qy}.
Given their central role in cosmology, FLRW space embeddings have been extensively studied in this higher-dimensional context.
However, they remain mostly grounded to $\setR^{n+1}$ and often involve cumbersome formulas regarding their additional coordinate.\cite{Rosen:1965,Gulamov:2011ux,Akbar:2017vja}
These previous embeddings paid no particular attention to the conformal symmetry of FLRW spaces.
In the ambient space $\setR^{n+2}$, minding the conformal symmetry, we found explicit and remarkably simple embedding formulas for each type of FLRW spaces.

Finally, having the relationship between intrinsic and ambient quantities along with explicit embedding formulas, we can apply this entire framework to physically relevant fields.
We then return to the study of two-point functions of conformally invariant fields and in particular to their expression in FLRW spaces.
First, we consider the massless conformally coupled scalar field, for which we recover known results straightforwardly and in a nice way. 
Second, we address Maxwell's field, which is conformally invariant in $n=4$ dimensions.
This case is more involved due to its gauge invariance.
Ambient results concerning this field are known already.\cite{Binegar1982Mink,Binegar1982AdS,Bayen:1984dt,Huguet:2006fe,Huguet:2008js,Faci2009,queva:tel-00503186}
Now, using our ambient-space construction combined with the new embedding formulas, we extend them to FLRW spaces.
In particular, we present a remarkably simple expression [see Eq.~\eqref{eq:<aa>}] in the ambient space from which one can derive the two-point functions for all conformally flat geometries, including FLRW spaces.
Furthermore, this general point of view in $\setR^{6}$ allows us to gain a clearer understanding on the terms contributing to the photon propagator.
Thanks to this understanding, we simplify further these two-point functions obtaining new closed-form expressions that apply to general FLRW spacetimes, thereby, we hope, adding to the recent impressive studies\cite{Domazet:2024dil,Glavan:2025iuw} on this topic.\footnote{See also Ref.~\onlinecite{Cotaescu:2021tza} in spatially flat FLRW spacetimes.}

The paper is organized as follows.
In Sec.~\ref{sec:Geometric_Setting} we expose the geometric setting in $\setR^{n+2}$ in which the physical spacetime $X_f$ is realized as the intersection of an hypersurface with the null cone of $\setR^{n+2}$.
We show that locally conformally flat spaces can be embedded in such a manner and how the conformal group $SO(2,n)$ acts then.
In Sec.~\ref{exemples} we obtain explicit examples of embedding formulas.
First, we expose the strategy to find new embeddings from the known embeddings of Minkowski and (A)dS spaces.
Then, applying this strategy, we obtain new and simple embeddings for arbitrary FLRW spaces in $\setR^{n+2}$.
At the end of the section, we discuss various bibliographic resources pertaining to the embeddings of FLRW spaces.
In Sec.~\ref{sec:Correspondence} we address the geometric correspondence between ambient and intrinsic differential forms.
In particular, we establish the bijection between differential forms on $X_f$ and the, to be defined, strongly transverse ambient differential forms.
In Sec.~\ref{sec:Differential} we study the restriction of ambient differential operators on $p$-forms to the embedded manifold, on the one hand, and the extension of differential operators out of $X_f$ to the ambient space, on the other hand.
In Sec.~\ref{sec:Curvature}, from the defining function $f$, we obtain the Riemann tensor, the Ricci tensor, and the scalar curvature of $X_f$.
Then, knowing these tensors and the restriction formula for the Laplace--de Rham operator, thanks to the Weitzenböck formula, we obtain the restriction formula for the Laplace--Beltrami operator on $p$-forms.
In Sec.~\ref{sec:Two-point_functions}, applying methods and formulas just obtained, we derive new expressions of the two-point functions of the massless conformally coupled scalar field and of Maxwell's field on FLRW spaces.
In particular, with our ambient viewpoint and the explicit relationship between different expressions of the photon propagator, we are able to simplify the presentation of Maxwell's field two-point function.
We have put an emphasis on producing explicit formulas that could be reused in other work.
We conclude in Sec.~\ref{sec:Conclusion}.
Four appendixes complement the main body of the article.
In App.~\ref{sec:Useful_formulas} we recall necessary formulas for our venture.
First, we set our notations and conventions regarding the differential calculus.
To these, we additionally collect properties of the Schouten--Nijenhuis bracket on forms and of the Kulkarni--Nomizu product, which is helpful when dealing with curvature.
In App.~\ref{app:Additional_Embedding} we expose the logic underlying the embeddings of FLRW spaces and in particular how to look for the simplest of them.
This involves properties specific to the de Sitter space that we take full advantage of.
In App.~\ref{sec:IsoFLRW}, for completeness, we retrieve the isometries of FLRW spaces in ambient space.
Finally, App.~\ref{sec:Proofs} collects the proofs whose length would impede the reading of the main body of the article.

\section{Geometric setting of the embeddings and their relationship with $SO(2,n)$}
\label{sec:Geometric_Setting}

\subsection{Main definitions}

The ambient space $\setR^{n+2}$ is endowed with its pseudo-Euclidean metric $\eta = \mathrm{diag}(+ -  \cdots - - +)$,
in the natural basis $\dr_\alpha = \frac{\dr\ }{\dr y^\alpha}$ with the $\{y^\alpha\}$ the Cartesian coordinates and $\alpha = 0, 1, ..., n-1, n, n+1$, as is the case for the indices $\beta, \gamma, ...$
Moreover, Greek letters in the middle of the alphabet, such as $\mu$, $\nu$, $\rho$,  range from $0$ to $n-1$ and refer to the spacetime.
Finally, Latin letters such as $i,j,k,m,n$ range from $1$ to $n-1$.
In the appendixes we will sometimes use the indices $a, b, c\ldots$ which can take the place of either one of the preceding sets.

\begin{definition}
    Let $c(y) = \frac{1}{2} y_\alpha y^\alpha$, then the null cone $\calC$ is defined as
    \begin{equation*}
        \calC = \{y\in\setR^{n+2}\ |\ c(y) = \tfrac{1}{2}y_\alpha y^\alpha = 0\}.
    \end{equation*}
    Associated to $c$ we define the vector field
    \begin{equation*}
        D = \sharp_\eta dc = y^\alpha\dr_\alpha,
    \end{equation*}
    known as the (ambient) dilation operator.
    The squared norm of $D$ reduces to $\mathbf{D}^2 = (\sharp_\eta dc)^2 = \eta(D,D) = y_\alpha y^\alpha = 2c(y)$.
\end{definition}

\begin{remark}
    Since various metrics will occur simultaneously we chose to be very explicit with respect to metric dependent quantities by subscripting them.
    This is the case with the musical operators $\sharp_\eta$ and $\flat_\eta$ referring, here, to the ambient metric $\eta$.
\end{remark}

\begin{definition}
    Let $f$ be a homogeneous function of degree $1$ in $\setR^{n+2}$, we define the hypersurface
    \begin{equation*}
        \calP_f = \{y\in\setR^{n+2}\ |\ f(y) = 1\}.
    \end{equation*}
    Associated to $f$ we define the vector field
    \begin{equation*}
        F = \sharp_\eta df = (\dr^\alpha f)\dr_\alpha,
    \end{equation*}
    with its squared norm  $\Fd = (\sharp_\eta df)^2 = \eta(F,F) = (\dr_\alpha f)(\dr^\alpha f)$.
\end{definition}

\begin{definition}\label{def:Xf}
    In $\setR^{n+2}$ we define the $n$-dimensional submanifold
    \begin{equation*}
        X_f = \calC\cap\calP_f.
    \end{equation*}
    We denote then the canonical injection $m_f : X_f\to\setR^{n+2}$ and $\eta_f = m_f{}^*\eta$ the induced metric on $X_f$.
\end{definition}

\begin{figure}
    \centering
    \includegraphics{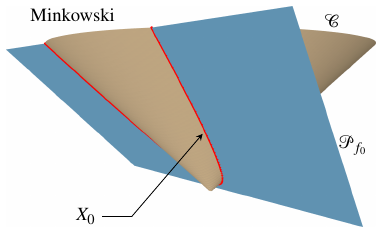}
    \hfill
    \includegraphics{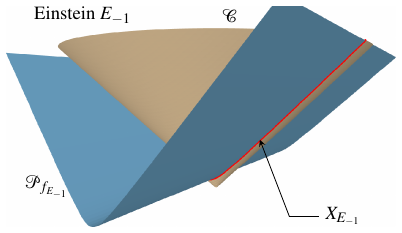}
    \caption{\label{fig:Xf_intersection}
        Submanifold $X_f$, red curve, viewed as the intersection of an hypersurface $\calP_f$, blue surface, and the null cone $\calC$ in $\setR^{n+2}$.
        On the left-hand side the case corresponding to a Minkowski space.
        On the right-hand side the Einstein space with FLRW metric of type $k=-1$ as given in Eq.~\eqref{eq:f_Einstein}.
    }
\end{figure}

The submanifold $X_f$ is the main protagonist as it is the conformally flat spacetime (if nonempty) of interest, see Fig.~\ref{fig:Xf_intersection} for a schematic picture of the embedding.
The vector fields $D$ and $F$ operate (via the interior product or the Lie derivative) on fields in $\setR^{n+2}$ and can be used to enforce constraints related to the manifold $X_f$.\cite{Zapata:2017gqg}

\begin{remark}
    From their definition one gets the useful relations
    \begin{equation*}
        D(f) = F(c) = \eta(D,F) = f,\quad
        D(c) = 2c,\quad
        D(\Fd) = 0,\quad
        [D,F] = -F,\quad
        F(f) = \Fd.
    \end{equation*}
\end{remark}

\begin{notation}
    The pull-back $m_f{}^*$ will play a central role in our study since it allows us to connect the differential forms of the ambient space with those of $X_f$. This link consists of a double restriction.
    First, on the application point $y$ which must belong to $X_f$ and we will note $\alpha_{|X_f}(y)=\alpha(y)$ as soon as $y\in X_f$. 
    Second, we restrict the vectors to which the form applies and we will note $\alpha_f=m_f{}^*\alpha$, that is to say $\alpha_f(y)(u_1,\ldots,u_a)=(m_f{}^*\alpha)(y)(u_1,\ldots,u_a)=\alpha(y)(u_1,\ldots,u_a)$ as soon as $y\in X_f$ and $u_1,\ldots,u_a\in T_y(X_f)$.
\end{notation}

\subsection{Main properties fulfilled by the embedded spaces $X_f$}

In the following we show that the metric manifold $(X_f, \eta_f)$ encompasses a wide array of physically relevant spacetimes, including FLRW spaces.
We first single out, among the $X_f$, the spaces of constant curvatures, namely de Sitter, anti--de Sitter, and Minkowski spaces.
Then we address the case of a generic $X_f$ proving that it is well-defined and that we recover, at least locally, conformally flat spaces.

\subsubsection{The special case of maximally symmetric spaces}

Maximally symmetric spaces appear throughout this work, their importance calls for a notation specific to them.

\begin{notation}
    Let us note $(\Sigma_\kappa, g_\kappa)$, with the constant $\kappa = H^2, 0, -H^2$, the maximally symmetric spaces of dimension $n$ and of constant curvature $\R_\kappa = -n(n-1)\kappa$.
    Those spaces are, respectively, identified to de Sitter (dS), Minkowski, and anti--de Sitter (AdS) spaces. 
    In the sequel, we will use the abbreviation AdSM to denote these three spaces.
\end{notation}

As an example from an earlier work\cite{Huguet:2023xmn} by setting $f(y) = H y^{n+1}$ the space $X_f$ identifies with a de Sitter space $(\Sigma_{\sss H^2},g_{\sss H^2})$.
Similarly by setting $f(y) = H y^n$ one obtains an AdS space.
Finally, by setting $f(y) = y^n + y^{n+1}$ one obtains a Minkowski space.
Those three functions $f$ and their associated spacetimes are rather general according to the following property.

\begin{property}\label{prop: AdSM}
    Let $f$ be a homogeneous polynomial of degree $1$, then $\Fd = \kappa$ is a constant and then
    \begin{equation*}
        (X_f,\eta_f) \simeq (\Sigma_\kappa,g_\kappa),
    \end{equation*}
    with $\simeq$ meaning that both spaces are diffeomorphically isometric to each other.
\end{property}
\begin{proof}
    The function $f$ can be written as $f(y) = A_\alpha y^\alpha$ with $A$ a constant vector, $\calP_f$ is then an hyperplane in $\setR^{n+2}$.
    Since $F = A^\alpha\dr_\alpha$ is a constant vector, $\Fd = A^\alpha A_\alpha = \kappa$ is also a constant.
    Then, depending on the sign of $\kappa$, it suffices to recognize that with an appropriate choice of frame in $\setR^{n+2}$ one falls into one of the three previous examples, reducing to an AdSM space.
\end{proof}

\begin{remark}
    For a smooth deformation between dS, Minkowski and AdS spaces one can use\cite{Huguet:2006fe}
    \begin{equation*}
        f(y) = \tfrac{1}{2}(1+\xi)y^{n+1} + \tfrac{1}{2}(1-\xi)y^n,
    \end{equation*}
    with $\xi = H^2 > 0$ for a dS space, resp. $\xi = 0$ for Minkowski space and $\xi = - H^2 < 0$ for an AdS space (see also Ref.~\onlinecite{Guo:2007vyt} for a similar approach).
\end{remark}

\subsubsection{The general case for $X_f$}

Beyond the special cases $(\Sigma_\kappa,g_\kappa)$ we want to generally characterize $X_f$.
This is first investigated by checking that the induced metric is nondegenerate (remember that, since it is defined on the null cone, caution is needed).
Then, we establish that $X_f$ spaces are conformally AdSM spaces and that conversely a globally conformally AdSM space (see definition below) can be embedded as an $X_f$ in $\setR^{n+2}$.
By construction, the (conformal) group $SO(2,n)$ acts on $X_f$ by conformal transformations.

\begin{property}\label{prop:TR6}
    The induced metric $\eta_f = m_f{}^*\eta$ is non-degenerated on $X_f$.
\end{property}
\begin{proof}
    This property edges upon the fact that at $y\in X_f\subset\setR^{n+2}$ one has
    \begin{equation}\label{eq:Split_TR6}
        T_y(\setR^{n+2}) = T_y(X_f)\stackrel{\perp}{\oplus}\mathrm{span}(F,D),
    \end{equation}
    with $T_y(\setR^{n+2})\simeq \setR^{n+2}$.
    First note that, at $y\in X_f$, $F$ and $D$ are two linearly independent vectors since $\eta(F,D)  = \langle df,D\rangle = D(f) = f = 1$ while $\eta(D,D) = 2c(y) =0$, thus they span a two-dimensional vector space.
    Then, vectors spanned by $F$ and $D$ are orthogonal to $T_y(X_f)$, with respect to $\eta$, by construction.
    Now, consider the vector $V = \alpha F + \beta D\in (T_y(X_f)\cap\mathrm{span}(F,D))$, with $\alpha$ and $\beta$ two reals, then since $\eta(V,D) = \beta = 0$ and $\eta(V,F) = \alpha = 0$ one concludes that the intersection is reduced to the null vector.
    This establishes Eq.~\eqref{eq:Split_TR6}.
    
    As a consequence there are no isotropic vectors in $T_y(X_f)$ that are orthogonal to $T_y(X_f)$ and $\eta_f(u,v) = m_f{}^*\eta(u,v)$, for $u,v\in T_y(X_f)$, is nondegenerate.
\end{proof}

From Ref.~\onlinecite{Huguet:2023xmn} we recall the following important property, details (in particular of the proof) are to be found in this previous article. 
Notations have been adapted to fit in the present article.

\begin{property}[Ref.~\onlinecite{Huguet:2023xmn}, Theorem~1]\label{prop:TheoremPRD}
    Let $f_1$ and $f_2$ be two homogeneous functions of degree $1$ in $\setR^{n+2}$, let $\phi = \ln(f_1/f_2)$ be then an homogeneous function of degree zero.
    Let $\Lambda : y\mapsto e^\phi y$ be the corresponding dilation in the ambient space.
    Then $\Lambda$ induces a diffeomorphic isometry
    \begin{equation*}
        (X_{f_1}, e^{2\phi}\eta_{f_1})\simeq(X_{f_2}, \eta_{f_2}).
    \end{equation*}
    Moreover, the elements of the linear group $SO(2,n)$ act on $X_{f_1}$ and $X_{f_2}$ as conformal transformations.
\end{property}

\begin{figure}
    \centering
    \includegraphics{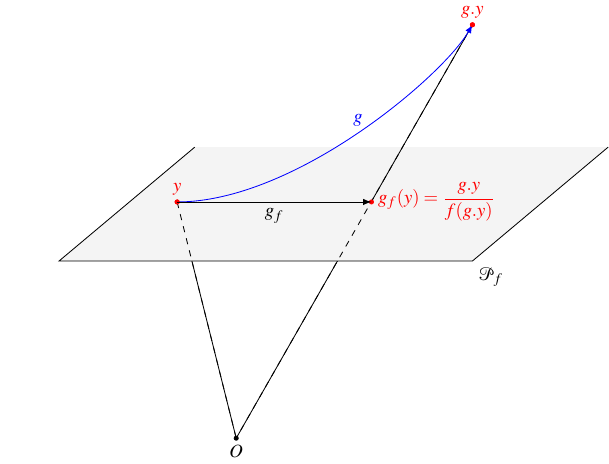}
    \caption{\label{fig:Action_SO24}
        Schematic depiction of the action of $SO(2,n)$ according to Eq.~\eqref{eq:g_f(y)} on points of $X_f\subset\calP_f$.
        If $g\in SO(2,n)$ is an isometry of $X_f$ the point remains on $\calP_f$ and $g_f(y) = g.y$.
        Otherwise, if $g$ is not an isometry of $X_f$, $g.y$ is brought back on $\calP_f$ by dilation along the half-lines of the null cone $\calC$ emanating from the origin $O$ in $\setR^{n+2}$.
    }
\end{figure}

Precisely, for $y\in X_f$ and $g\in SO(2,n)$, one has
\begin{equation}
    \label{eq:g_f(y)}
    g_f(y) = \frac{g.y}{f(g.y)}\in X_f,
\end{equation}
with $g.y$ the natural action of $SO(2,n)$ in $\setR^{n+2}$, see Fig.~\ref{fig:Action_SO24}.
This is a conformal transformation since
\begin{equation*}
    \eta(g_{f*}(u),g_{f*}(v))
    = \frac{1}{f^2(g.y)}\eta(u,v),\quad
    u,v\in T_y(X_f).
\end{equation*}

\begin{definition}
    A metric space $(M,g)$ will be named globally conformally AdSM if there exists a function $\phi$ on $\Sigma_\kappa$ and an isometric
    conformal diffeomorphism $\Xi: (M,g)\to(\Sigma_\kappa, e^{2\phi}g_\kappa)$.
    This, in turn, induces a conformal diffeomorphism between $(M,g)$ and $(\Sigma_\kappa,g_\kappa)$.
\end{definition}

\begin{property}\label{prop:BigProp}
    The submanifold $X_f$ with its induced metric $\eta_f$ is either empty,
    or a conformally flat space.
    Conversely, let $(M,g)$ be a $n$-dimensional globally conformally AdSM space, then there exists an homogeneous function $f$ of degree $1$ in $\setR^{n+2}$ such that $M$ and $X_f$ are isometric.
\end{property}
\begin{proof}
    First, let us show that $X_f$, if not empty, is conformally flat.
    Let $y_0$ be a given point of $X_f$ and $f_0$ be an homogeneous polynomial of degree $1$ such that $f_0(y_0)=1$, for instance $f_0(y)=F_\alpha(y_0)y^\alpha$.  
    Then, in a neighborhood of $y_0$ we define the homogeneous function of degree zero $\phi_0 = \ln(f_0/f)$, which is well defined since $f(y)\approx 1\approx f_0(y)$ and with $\phi_0(y_0) = \ln 1 = 0$.
    Then, according to Prop.~\ref{prop:TheoremPRD}, the map $\Lambda_0 : y\mapsto e^{\phi_0}y$ induces the isometry $(\Sigma_{k_0}, e^{2\phi_0}g_{k_0}) \simeq (X_f, \eta_f)$, thus $X_f$ is 
    locally conformally flat since $\Sigma_\kappa$ is.

    Second, let us show the converse result, that is that a globally conformally AdSM space $M$ can be isometrically embedded as an $X_f$ in $\setR^{n+2}$.
    Since $M$ is globally conformally AdSM there exists $\Xi$ mapping it to $(\Sigma_\kappa, e^{2\phi}g_\kappa)$ with $\phi$ defined on $\Sigma_\kappa$.
    The function $\phi$ is extended from $\Sigma_\kappa$ to $\calC\subset\setR^{n+2}$ by asking it to obey $D(\phi) = 0$.
    Then, according to Prop.~\ref{prop:TheoremPRD}, the map $\Lambda : y\mapsto e^{\phi}y$ induces the isometry $(\Sigma_\kappa, e^{2\phi}g_\kappa) \simeq (X_f, \eta_f)$ with then $f = e^{-\phi} f_\kappa$ and $f_\kappa$ the homogeneous polynomial of degree $1$ defining $\Sigma_\kappa$.
    Then the map $\Lambda\circ\Xi$ performs the embedding of $M$ as $X_f$ in $\setR^{n+2}$.
\end{proof}

With this property we recover in a simple setting known facts about conformally flat manifolds as hypersurfaces of a null cone of a $(n+2)$-dimensional Lorentz space.\cite{Brinkmann_1923,Asperti_Dajczer_1989}

\section{Explicit examples of embeddings in $\setR^{n+2}$ (including FLRW spaces)}
\label{exemples}

\subsection{Methods of embeddings by leveraging Prop.~\ref{prop:TheoremPRD}}

Let $f_\kappa$ be the homogeneous polynomial that defines the constant curvature space $X_{f_\kappa}= X_\kappa \simeq~\Sigma_\kappa$ (see Prop.~\ref{prop: AdSM}) and let $\phi$ be a function on $X_\kappa$.
We note $\tilde\phi$ an extension of $\phi$ to the ambient space which is homogeneous of degree 0.
We then set
\begin{equation}
    \label{Eq : exemple}
    f=e^{-\tilde\phi} f_\kappa.
\end{equation}
The homogeneity of $\tilde\phi$ means that it can be seen, indifferently, as a function on $X_\kappa$ or on $X_f$.
The direct application of Prop.~\ref{prop:TheoremPRD} shows that $X_f$ is globally conformally AdSM with $e^{2\phi}$ as the Weyl factor: $\eta_f=e^{2\phi}g_\kappa$ where $g_\kappa$ is the AdSM metric.
We recall that this metric rescaling is induced by the rescaling of the coordinates on $X_\kappa$ as $\Lambda : y_\kappa\mapsto e^\phi y_\kappa$ . %
We often say that $X_\kappa$ is the \emph{base space} via which the embedding of $X_f$ is performed.
The important point is that any coordinate system of $X_\kappa$ provides a coordinate system for $X_f$ in which the metric is written in a simple way.

\begin{remark}
    Every conformally flat space can be put,  locally,  in the form \eqref{Eq : exemple} and every globally conformally AdSM space can be put, globally, in this form.
    (See the proof of Prop.~\ref{prop:BigProp}.)
\end{remark}

\subsection{Parametrization of AdSM spaces}

As a first step, we begin with obtaining a parametrization of 
the AdSM space $X_\kappa$ for the three possible cases: $\kappa=0$, $H^2$, $-H^2$ successively.

For $\kappa=0$, $X_0$ is a Minkowski space defined by $f_0(y)=y^n+y^{n+1}$, a global parametrization is given by
\begin{equation}
    \label{eq:Coord_Mink}
    \left\{
    \begin{aligned}
        &y^\mu=\xi^\mu,\\
        &y^n=\tfrac{1}{2}(1+\xi_\mu \xi^\mu),\\
        &y^{n+1}=\tfrac{1}{2}(1-\xi_\mu \xi^\mu),
    \end{aligned}
    \right.
\end{equation}
with $\xi\in\setR^n$.
Furthermore,
\begin{equation*}
    \eta_{0}=\eta_{\alpha\beta}dy^\alpha \otimes dy^\beta=\eta_{\mu\nu}d\xi^\mu \otimes d\xi^\nu.
\end{equation*}

For $\kappa\neq0$ there is no global map. 
However, local maps can be obtained.
For $\kappa=H^2$ we obtain a parametrization of the de Sitter space $X_{\sss H^2}$, defined by $f_{\sss H^2}(y)=Hy^{n+1}$, as follows.
Let ${\cal U}=\{ \xi\in\setR^4\ |\ H^2\xi_\mu \xi^\mu>-1\}$ and %
\begin{equation*}
\left\{
    \begin{aligned}
    &y^\mu=\xi^\mu,\\
    &y^n=\sqrt{\xi_\mu \xi^\mu+H^{-2}},\\
    &y^{n+1}=H^{-1},
    \end{aligned}
\right.
\end{equation*}
which parametrizes half ($y^n>0$) of $X_{\sss H^2}$. Then
\begin{align*}
\eta_{\sss H^2}&=\eta_{\alpha\beta}dy^\alpha \otimes dy^\beta
=\left(\eta_{\mu\nu}
-\frac{\xi_\mu \xi_\nu}{\xi_\lambda \xi^\lambda+H^{-2}}\right)d\xi^\mu\otimes d\xi^\nu,
\end{align*}
where we recognize the metric on de Sitter space in its ``natural'' embedding coordinates, see, e.g., Sec.~13.3 of Ref.~\onlinecite{Weinberg:1972kfs}.
Finally, for $\kappa=-H^2$ the anti--de~Sitter space $X_{\sss -H^2}$, defined by $f_{\sss -H^2}(y)=Hy^n$, we obtain the parametrization
\begin{equation*}
\left\{
\begin{aligned}
    &y^\mu=\xi^\mu,\\
    &y^{n}=H^{-1},\\
    &y^{n+1}=\sqrt{-\xi_\mu \xi^\mu+H^{-2}},
\end{aligned}
\right.
\end{equation*}
for $\xi\in{\cal V}$ where ${\cal V}=\{ \xi\in\setR^4\ |\ H^2\xi_\mu \xi^\mu<1\}$. Then
\begin{equation*}
    \eta_{\sss -H^2}
    = \eta_{\alpha\beta} dy^\alpha\otimes dy^\beta
    =\left(\eta_{\mu\nu}
    -\frac{\xi_\mu \xi_\nu}{\xi_\lambda \xi^\lambda-H^{-2}}\right)d\xi^\mu\otimes d\xi^\nu.
\end{equation*}

\subsection{Parametrization of conformally AdSM spaces}

A parametrization of the spaces with base space $X_\kappa$ is now easily obtained, depending on the value of $\kappa$.
In the following we set $\psi(\xi)=\tilde\phi(y(\xi))$.

For $\kappa=0$, i.e. for Minkowski space as base space, coordinates on $X_f$ [Eq.~\eqref{Eq : exemple}] are given by
\begin{equation}
    \label{eq:CAdSM-Mink}
    \left\{
    \begin{aligned}
        &y^\mu=e^{\psi(\xi)}\xi^\mu,\\
        &y^n=\tfrac{1}{2}e^{\psi(\xi)}(1+\xi_\mu \xi^\mu),\\
        &y^{n+1}=\tfrac{1}{2}e^{\psi(\xi)}(1-\xi_\mu \xi^\mu),
    \end{aligned}
    \right.
\end{equation}
with $\xi\in\setR^n$, and 
\begin{equation*}
    \eta_{f}=e^{2\psi(\xi)}\eta_{\mu\nu}d\xi^\mu \otimes d\xi^\nu.%
\end{equation*}

For $\kappa= -H^2$ and $H^2$, coordinates on $X_f$ [Eq.~\eqref{Eq : exemple}] are given by
\begin{equation*}
\kappa=-H^2 : \left\{
\begin{aligned}
&y^\mu=e^{\psi(\xi)}\xi^\mu,\\
&y^{n}=e^{\psi(\xi)}H^{-1},\\
&y^{n+1}=e^{\psi(\xi)}\sqrt{-\xi_\mu \xi^\mu+H^{-2}},
\end{aligned}
\right.
\quad\kappa=H^2 :
\left\{
\begin{aligned}
    &y^\mu=e^{\psi(\xi)}\xi^\mu,\\
    &y^n=e^{\psi(\xi)}\sqrt{\xi_\mu \xi^\mu+H^{-2}},\\
    &y^{n+1}=e^{\psi(\xi)}H^{-1},
\end{aligned}
\right.
\end{equation*}
where, in the first case ($\kappa=-H^2$, AdS as base space) we have $\xi\in{\cal V}=\{\xi\ |\ H^2\xi_\mu\xi^\mu<1\}$ and 
\begin{equation*}
    \eta_{f}
    =e^{2\psi(\xi)}\left(\eta_{\mu\nu}
    -\frac{\xi_\mu \xi_\nu}
        {\xi_\lambda \xi^\lambda-H^{-2}}\right)d\xi^\mu\otimes d\xi^\nu,
\end{equation*}
while, in the second case
($\kappa=H^2$, dS as base space) we have $\xi\in{\cal U}=\{\xi\ |\ H^2\xi_\mu\xi^\mu > -1\}$ and 
\begin{equation*}
    \eta_{f}
    =e^{2\psi(\xi)}\left(\eta_{\mu\nu}
    -\frac{\xi_\mu \xi_\nu}
        {\xi_\lambda \xi^\lambda+H^{-2}}\right)d\xi^\mu\otimes d\xi^\nu.
\end{equation*}

\subsection{The special case of FLRW spaces}

The purpose of this section is to provide a parametrization, as simple as possible,  of the FLRW spaces in ambient space and to write the corresponding defining functions $f_k$.

FLRW spaces fulfill the homogeneity and isotropy hypotheses of space\cite{Robertson_1935,Walker_1937} for which the metric can be written as
\begin{equation}
    \label{eq:EtaRW}
    \eta_{\sss{FLRW}}
    = a^2(t)\left[dt\otimes dt - \frac{1}{1-kr^2}dr\otimes dr - r^2g_{\mathbb{S}^{n-2}}\right],
\end{equation}
with $t$ the conformal time,\footnote{The conformal time, most usually, is written as $\eta$. A choice we are unable to make, $\eta$ being the metric of ambient
space in this work.}
$a$ the conformal scale factor, $k\in\{-1,0,+1\}$ depending on the choice of curvature of the $n-1$ space at constant conformal time, and $g_{\mathbb{S}^{n-2}}$ is the ordinary metric on the sphere $\mathbb{S}^{n-2}$. 
We say that a space equipped with the metric defined in Eq.~\eqref{eq:EtaRW} is a FLRW space of type $k$.
Often we use the variable $\chi$, which is related to the radial coordinate $r$ of the isotropic form of the spatial metrics by $r = \sin(\chi)$ for $k=+1$, $r= \sinh(\chi)$ for $k=-1$, and $r = \chi$ for $k=0$.
Note that in the above we simply consider Eq.~\eqref{eq:EtaRW} as defining FLRW spaces, the scale factor $a(t)$ is left unspecified.

We derived formulas that provide embeddings of FLRW spaces into $\setR^{n+2}$ in the simplest manner possible, as captured in the following property.
\begin{property}
    \label{prop:SimpleEmb}
    For a given arbitrary scale factor $a$, and for $\omega^i$ ambient coordinates on $\mathbb{S}^{n-2}$ [i.e., $\sum_i (\omega^i)^2 = 1$], the following formulas
    \begin{align}
        \label{eq:SimpleEmb}
        &\left\{
            \begin{aligned}
                &y^0 = a(t)\cosh(\chi), \\
                &y^i = a(t)\sinh(\chi)\omega^i,\\
                &y^n = a(t)\cosh(t),\\
                &y^{n+1} = a(t)\sinh(t),
            \end{aligned}\right.
        &&\left\{
            \begin{aligned}
                &y^0 = a(t)t,\\
                &y^i = a(t)\chi \omega^i,\\
                &y^n=\tfrac{1}{2}a(t)(1+t^2 -\chi^2),\\
                &y^{n+1}=\tfrac{1}{2}a(t)(1-t^2 + \chi^2),
            \end{aligned}\right.
        &&\left\{
            \begin{aligned}
            &y^0=a(t)\cos(t),\\
            &y^i=a(t)\sin(\chi)\omega^i,\\
            &y^n=a(t)\cos(\chi),\\
            &y^{n+1}=a(t)\sin(t),
            \end{aligned}
        \right.
    \end{align}
    yield an embedding in $\setR^{n+2}$ of FLRW spaces of type $k=-1$, $k=0$ and $k=+1$, with $\chi\in\setR$ in the first two cases and $\chi\in\setR/2\pi\setZ$ in the last case. 
\end{property}

\begin{figure}
    \centering
    \includegraphics{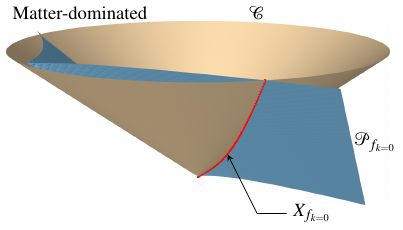}
        \hfill
    \includegraphics{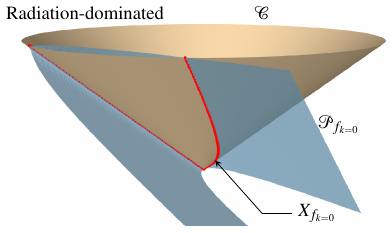}
    \caption{\label{fig:Matter-Radiation}
        Depiction in ambient space of a matter-dominated FLRW space of type $k=0$, red curve on the left, with $a(\tau) \propto \tau^{2/3}$, that is $a(t) \propto t^2$ in conformal time.
        On the right-hand side, red curve also, a radiation-dominated FLRW space of type $k=0$ with $a(\tau) \propto \tau^{1/2}$, that is $a(t) \propto t$ in conformal time.
        The embeddings are obtained using directly the defining functions of Eq.~\eqref{eq:f_FLRW}.
    }
\end{figure}

It is straightforward to verify through calculations that this statement holds true.
The logic underlying the calculations resulting in Prop.~\ref{prop:SimpleEmb} and Eq.~\eqref{eq:SimpleEmb}, exploiting properties specific to dS and Minkowski spaces, is reported in App.~\ref{app:Additional_Embedding}.
In Fig.~\ref{fig:Matter-Radiation} we plotted in ambient space a matter-dominated and radiation-dominated FLRW spaces commonly used in cosmology.

\begin{remark}
    The formula for the $k=0$ FLRW case in Cartesian coordinates $\xi$, setting $\xi^0 = t$ and $\xi^i = \chi\omega^i$, reads
    \begin{equation}
    \label{eq:SimpleEmbk=0}
        \left\{
        \begin{aligned}
            &y^\mu=a(\xi^0)\xi^\mu,\\
            &y^n=\tfrac{1}{2}a(\xi^0)(1+\xi_\mu \xi^\mu),\\
            &y^{n+1}=\tfrac{1}{2}a(\xi^0)(1-\xi_\mu \xi^\mu),
        \end{aligned}
        \right.
    \end{equation}
    which can be seen directly from Eq.~\eqref{eq:CAdSM-Mink} with $e^{\psi(\xi)} = a(\xi^0)$.
\end{remark}

The conformal group $SO(2,n)$ acts on FLRW spaces.
Among the $\tfrac{1}{2}(n+1)(n+2)$ generators of the conformal group at least $\tfrac{1}{2}(n-1)n$ are Killing vectors of the FLRW metric generating the subgroup $SO(1,n-1)$, $E(n-1)$, and $SO(n)$ of $SO(2,n)$ depending on $k$.
Note that these (minimal) isometries of the FLRW metrics can directly be read off of the embedding formulas Eq.~\eqref{eq:SimpleEmb}.
Here, from the ambient space point of view in $\setR^{n+2}$, the isometries of $X_f$ are  found as combinations of the generators of $o(2,n)$ which commute with the defining function $f$.
In App.~\ref{sec:IsoFLRW} we recover these isometries and the exceptional additional ones, in the ambient space $\setR^{n+2}$, using the defining functions $f_k$ displayed in Eq.~\eqref{eq:f_FLRW}.

Within $\setR^{n+2}$, and according to Eq.~\eqref{Eq : exemple}, one can obtain each type of FLRW space by using the defining function $f_k$, not to be confused with $f_\kappa$ appearing in Eq.~\eqref{Eq : exemple} defining AdSM spaces, as
\begin{equation}
    \label{eq:f_FLRW}
    f_k(y) = \begin{cases}
        \dfrac{1}{a(\tanh^{-1}(y^{n+1}/y^n))}
        \sqrt{(y^n)^2 - (y^{n+1})^2} &\text{for $k=-1$,}\\
        \dfrac{1}{a(y^0/(y^n+y^{n+1}))}(y^n + y^{n+1}) &\text{for $k=0$,}\\
        \dfrac{1}{a(\tan^{-1}(y^{n+1}/y^0))}
            \sqrt{(y^0)^2 + (y^{n+1})^2} &\text{for $k=+1$.}
    \end{cases}
\end{equation}
These defining functions can be recovered thanks to Eq.~\eqref{Eq : exemple} with the defining function of a dS space $f_\kappa(y) = Hy^{n+1}$ and the conformal factors $\Omega = e^{\tilde\phi}$ found in Eq.~\eqref{eq:Ak_dS} of App.~\ref{app:Additional_Embedding}.
That is, on $X_{f_k} = \calC\cap\calP_{f_k}$ with $f_k$ as above, the induced metric $\eta_k$ is such that it can be turned into the FLRW metric~\eqref{eq:EtaRW} with the appropriate scale factor $a$ and of the desired $k$ type.
Evidently, the coordinate systems in Eq.~\eqref{eq:SimpleEmb} fulfill both $c(y) = y_\alpha y^\alpha/2 = 0$ and $f_k(y) = 1$.

\begin{remark}
    The simplest of those FLRW spaces are the Einstein spaces $E_k$ for which $a=1$.
    The defining functions for them are, from Eq.~\eqref{eq:f_FLRW}, then
    \begin{equation}
        \label{eq:f_Einstein}
        f_{\sss{E}_k}(y) = \begin{cases}
            \sqrt{(y^n)^2 - (y^{n+1})^2} &\text{for $k=-1$,}\\
            \sqrt{(y^0)^2 + (y^{n+1})^2} &\text{for $k=+1$,}
        \end{cases}
    \end{equation}
    with $k=0$ being nothing other than Minkowski space that we suppressed.
\end{remark}

The embeddings of FLRW spaces can be traced back to their initial investigation by Robertson.\cite{Robertson:1928,Robertson:1933zz}
Occasionally, they are independently rediscovered,\cite{Rosen:1965,Lachieze-Rey:2000gpy,Smolyakov:2008dr} unaware of earlier works.
The general wisdom\cite{Robertson:1933zz,Rosen:1965,Lachieze-Rey:2000gpy,Smolyakov:2008dr,Gulamov:2011ux} is that these embeddings require only one additional embedding dimension (class $1$ embeddings).
Then, often, a redundant coordinate is added in hopes of making the (ambient) metric flat, while matching the induced metric with that of the target FLRW spaces.
Frequently, this extra-coordinate value is obtained by an integral formula, where obtaining closed-form expressions appears challenging\cite{Rosen:1965,Gulamov:2011ux,Akbar:2017vja} (apart, maybe, in the simplest models).
Paston and Sheykin\cite{Paston:2013uia} took a different approach and used the isometries of FLRW spaces to build them in the five-dimensional flat embedding space.
Akbar\cite{Akbar:2017vja} noted that embeddings of FLRW spaces correspond to unit-speed parametrized curves in either two or three dimensions. Additionally, he pointed out the requirement for two extra embedding coordinates for certain unique types of FLRW spaces.
For depictions of such embeddings, for different kind of cosmological FLRW spaces, in $\setR^3$ see Refs.~\onlinecite{Lynden-Bell_1989,Lachieze-Rey:2000gpy,Wesson_2001,Seahra:2002ib}.

Throughout our literature review, we identified no embeddings of FLRW spaces as straightforward as those presented in Eq.~\eqref{eq:SimpleEmb}.
The closest examples we found are in brane-localized gravities in which Robertson-Walker spaces were embedded in AdS$_{n+1}$ spaces,\cite{Mannheim:2000zy} which themselves were embedded in $\setR^{n+2}$ (see Chap.~10 of Ref.~\onlinecite{Mannheim_2005}).
However, as they are realized on the hyperboloid of $\setR^{n+2}$, and not the null cone $\calC$, they lack the straightforward scaling symmetry of our embeddings.
More remotely the embeddings of Friedmann branes in five dimensional Schwarzschild--anti--de Sitter bulk\cite{Binetruy:1999hy,Bowcock:2000cq,Mukohyama:1999wi,Gergely:2001tn} could, in principle, themselves be embedded in $\setR^6$ using the embedding techniques devised in Ref.~\onlinecite{Giblin:2003xj}.

These embeddings have been used for investigations of the initial singularity in cosmological models,\cite{Lauro_1985,Lynden-Bell_1989,Rindler_2000,Seahra:2002ib}
    higher dimensional physics of Kaluza--Klein kind\cite{Ponce_de_Leon_1988,Wesson_1992,Wesson_1994,Bona:2018svb} (see in particular Refs.~\onlinecite{Wesson:2010qr,PoncedeLeon:2015pug} for such an approach)
    or of Regge--Teitelboim kind.\cite{Sheykin:2015nyi,Stern:2022uqc}
We hope that the embedding formulas presented here can assist in exploring cosmological phenomena such as in brane universes in six-dimensional flat spaces.
\cite{Gogberashvili:2002pp,Koley:2006bh,Gogberashvili:2017eie}

\section{The correspondence between the ambient forms $\Omega(\setR^{n+2})$ and those on the embedded spacetime $\Omega(X_f)$}
\label{sec:Correspondence}

This work deals, among other things, with differential forms and the differential calculus attached to them.
It is then of the utmost importance to relate as easily as possible the calculus on the embedded space with the one in the ambient space.
This is the correspondence that is worked out in this section, that is the one between $\Omega(X_f)$ and $\Omega(\setR^{n+2})$, both as a reduction and as an extension.

\subsection{A convenient basis of $T_y(\setR^{n+2})$ around $X_f$}

\begin{notation}
    Let $\{e_\mu\}$ be a free family defined in the vicinity of $X_f$ that becomes, at least locally, an orthonormal frame of $T_y(X_f)$ at each point $y\in X_f$.
    The vectors $F$ and $D$ span the orthogonal supplement in $T_y(\setR^{n+2})$ (see the proof of  Prop.~\ref{prop:TR6}).
    From those vectors, at $y\in X_f$, we define the orthonormal basis $\{e_\mu, e_n, e_{n+1}\}$ with
    \begin{align}
        &e_n = F -\tfrac{1}{2}(1+\Fd)D, \label{eq:en}\\
        &e_{n+1}= F +  \tfrac{1}{2}(1-\Fd)D \label{eq:en+1},
    \end{align}
    with $\eta(e_n,e_n) = - 1 = \eta_{nn}$ and $\eta(e_{n+1},e_{n+1}) = + 1 = \eta_{n+1\, n+1}$ for $y\in X_f$.
    In addition we have
    \begin{equation*}
        [e_n,e_{n+1}] = [F,D] = F.
    \end{equation*}
    Orientation is chosen such that for $\{e_\mu\}$ direct in $T_y(X_f)$ then $\{e_\mu,e_n, e_{n+1}\}$ is direct in $T_y(\setR^{n+2})$.
    By continuity they remain a basis in a neighborhood of $X_f$ in $\setR^{n+2}$, however orthonormality is lost.
\end{notation}

\begin{notation}
    To $\{e_\mu, e_n, e_{n+1}\}$ we associate the dual basis $\{e^\mu, e^n, e^{n+1}\}$, which is also a local orthonormal basis of $T_y^*(X_f)$ with then
    \begin{align*}
        &e^n =\eta^{nn}\flat_\eta e_n = -df + \tfrac{1}{2}(1+\Fd)dc,\\
        &e^{n+1} =\eta^{n+1\, n+1}\flat_\eta e_{n+1} = df + \tfrac{1}{2}(1-\Fd)dc.
    \end{align*}
\end{notation}
Directly from their definitions and App.~\ref{diff.geom.} one can notice the useful relations
\begin{align}
    &i_n i_{n+1} \equiv i_{e_n}i_{e_{n+1}}  = \iF\iD,\label{eq:InIn+1}\\
    &j^n j^{n+1} \equiv j^{e^n}j^{e^{n+1}} = -j^{df}j^{dc}.\label{eq:JnJn+1}
\end{align}

\subsection{Transverse and longitudinal ambient differential forms}

A differential form $\alpha\in\Omega^a(\setR^{n+2})$ around $X_f$ can be split in four parts depending whether $e^n$ or $e^{n+1}$ (or both, or none) appear in its expansion
\begin{align*}
    \alpha = &\frac{1}{a!}\alpha_{\mu_1\cdots\mu_a} e^{\mu_1}\wedge\cdots\wedge e^{\mu_a}\\
    &+\frac{1}{(a-1)!}\alpha_{\nu_1\cdots\nu_{a-1}n} e^{\nu_1}\wedge\cdots\wedge e^{\nu_{a-1}}\wedge e^n\\
    &+\frac{1}{(a-1)!}\alpha_{\nu_1\cdots\nu_{a-1}n+1} e^{\nu_1}\wedge\cdots\wedge e^{\nu_{a-1}}\wedge e^{n+1}\\
    &+\frac{1}{(a-2)!}\alpha_{\lambda_1\cdots\lambda_{a-2}nn+1} e^{\lambda_1}\wedge\cdots\wedge e^{\lambda_{a-2}}
        \wedge e^n\wedge e^{n+1}.
\end{align*}
The first term verifies
\begin{equation*}
    \frac{1}{a!} \alpha_{\mu_1\cdots\mu_a} e^{\mu_1}\wedge\cdots\wedge e^{\mu_a}
    = i_{n+1}i_nj^nj^{n+1}\alpha
    = \iF\iD j^{df}j^{dc}\alpha,
\end{equation*}
using Eqs.~\eqref{eq:InIn+1} and \eqref{eq:JnJn+1} for the second equality.
This leads us to the following definition.
\begin{definition}
    Setting
    \begin{equation*}
        T = \iF\iD j^{df}j^{dc},
    \end{equation*}
    then a differential form $\alpha\in\Omega^a(\setR^{n+2})$ is a transverse differential form with respect to $X_f$ iff $T\alpha=\alpha$ or, equivalently, iff $\iF\alpha = 0$ and $\iD\alpha = 0$.
    The set of transverse forms is noted $\Omega_{\rm t}(\setR^{n+2})$.
\end{definition}
    
\begin{remark}
    The identification $y=m(y)$ for $y\in X_f$ leads to the identification $m_*v=v$ for $v\in T(X_f) \subset T(\setR^{n+2})$ and then for $\lambda\in T^*(\setR^{n+2})$ and $v\in T(X_f)\subset T(\setR^{n+2})$ we have
$\langle\lambda,v\rangle
    =\langle\lambda,m_*v\rangle
    =\langle m^*\lambda,v\rangle$ and, as a consequence, using Eq.~\eqref{eq:Split_TR6}, $\lambda=m_f{}^*\lambda$ if and only if $\lambda$ is transverse.
    Then, since $m_f{}^*(e^n) = m_f{}^*(e^{n+1}) = 0$ and now $m_f{}^*(e^\mu) = e^\mu$, one obviously has
\begin{equation*}
    m_f{}^*\alpha=(T\alpha)_{|X_f}.   
\end{equation*}
Moreover we have
\begin{equation*}
    m_f{}^*\alpha=\alpha_{|X_f} \mbox{ iff }\alpha\in\Omega_{\rm t}(\setR^{n+2}).    
\end{equation*} 
Recall that we will sometimes use the shorter expression $\alpha_f=m_f{}^*\alpha$.
\end{remark} 

Any form $\alpha$ can be decomposed into the so-called transverse and longitudinal parts in the following way.
We set
\begin{equation*}
    L  = \Id-T
    = j^{df}\iD + j^{dc}\iF - \Fd j^{dc}\iD - j^{df}j^{dc}\iF\iD,
\end{equation*}
in which the last expression is obtained by putting $T = \iF\iD j^{df}j^{dc}$ in its normal form (the $i$'s on the right and the $j$'s on the left).
This leads us to the following definition.
\begin{definition}
A differential form $\alpha\in\Omega^a(\setR^{n+2})$ is said to be longitudinal with respect to $X_f$ if and only if $L\alpha=\alpha$.
The set of longitudinal forms is noted $\Omega_{\rm l}(\setR^{n+2})$.
\end{definition}

We then obtain the natural splitting into transverse and longitudinal parts in ambient space
\begin{equation*}
   \alpha= T\alpha+L\alpha.
\end{equation*}
Moreover, defining $m_f{}^l\alpha= (L\alpha)_{|X_f}$ we obtain at any point of $X_f$
\begin{equation}
    \alpha_{|X_f} = m_f{}^*\alpha + m_f{}^l\alpha, \label{Eq:trans-long}
\end{equation}
with
\begin{align*}
    &m_f{}^*\alpha = (T\alpha)_{|X_f} = (\iF\iD j^{df} j^{dc}\alpha)_{|X_f},\\
    &m_f{}^l\alpha = (L\alpha)_{|X_f} 
        =\left[(j^{dc}\iF - \Fd j^{dc}\iD + j^{df}\iF j^{dc}\iD)\alpha\right]_{|X_f}.
\end{align*}

\subsection{Strongly transverse ambient differential forms}
The mapping $m_f{}^*:\ \Omega_{\rm t}(\setR^{n+2}) \to\Omega(X_f)$ is a surjection, we seek a smaller initial set to make it a bijection.
Within the span of $F$ and $D$ we introduce a new vector field tangent to $\calP_f$.
\begin{definition}
    Let us define
    \begin{equation*}
        E_f =  fF-\Fd D,
    \end{equation*}
    which from direct computations fulfills the crucial properties
    \begin{equation*}
        \eta(E_f,F) = 0,\qquad
        [D,E_f] =0. %
    \end{equation*}
\end{definition}

This new vector is tangent to $\calP_f$ in particular and to $\{y\in\setR^{n+2}\ |\ f(y) = k\}$ in general.
Note that, on $X_f$, $\{D,E_f\}$ is the dual basis to $\{df,dc\}$.

We will now see how a differential form on spacetime can be extended to ambient space in such a way as to realize an isomorphism between the forms of ambient space and those of spacetime. It will appear that this isomorphism is well behaved with respect to differential operators.

\begin{property} 
    Let $\beta$ be a differential form on $X_f$, and $r\in\setR$, then, at least locally, there exists a unique differential form  $\hat{\beta}$ extending $\beta$ (i.e., $m_f{}^*\hat \beta=\beta)$ in a neighborhood of $X_f\subset\setR^{n+2}$ such that $(\LD-r\Id) \hat{\beta} = \L_{\sss E_f}\hat{\beta} = 0$.
    Moreover, $\hat\beta$ is transverse (i.e., $\iD\hat\beta=\iF\hat\beta=0$).
\end{property}
\begin{proof}
We begin with the definition of $\hat\beta$. 
Any point $p$ in a neighborhood of $X_f$ in $\calP_f$ can be written as $p=e^{t E_f}x_p$ with $x_p\in X_f$ and $t\in\setR$ a parameter.
We then set $\beta'(p) = (e^{-tE_f})^*\beta(x_p)=e^{-t\Lie_{E_f}}\beta(x_p)$.
Then $\beta'$ is extended as $\hat{\beta}$ by proceeding similarly from $\calP_f$ to $\setR^{n+2}$ using $\LD-r\Id$ instead of $\L_{E_f}$.
Namely, for $x= e^{tD}x_p$ we set $\hat\beta(x)= e^{-t(\LD-r\Id)}\beta'(x_p)$.
Since $[D,E_f] =0$, one has that $(\LD -r\Id)\hat{\beta} = \L_{\sss E_f}\hat{\beta} = 0$ as a result.
Uniqueness is a direct consequence of uniqueness of solutions of linear ordinary differential equations.
Finally,  we already know (by construction) that $\hat\beta_{|X_f}=\beta$ and that $\beta$ is transverse on $X_f$, as a consequence $m_f{}^*\hat\beta=\beta$.

    Moreover, the extended differential form remains transverse in $\Omega^b(\setR^{n+2})$.
    Indeed, since $[i_u,\Lie_v]=i_{[u,v]}$, $\iD$ and $\iF$ commute with $\LD-r\Id$ and $\Lie_{E_f}$ the result follows from the construction of $\hat\beta$ and the fact that $\iD\beta = 0$ and $\iF\beta = 0$.
\end{proof}
\begin{remark}
    The condition $(\LD-r\Id)\hat\beta = 0$ is equivalent to $\hat\beta$ being homogeneous of degree $r$.
\end{remark}
This leads to the following definition.
\begin{definition}
    A differential form $\alpha\in\Omega^a(\setR^{n+2})$ is a strongly transverse differential form of degree $r$ with respect to $X_f$ if $\iD\alpha = \iF\alpha = 0$ and $(\LD-r\Id)\alpha = \LF\alpha = 0$.
    The set of strongly transverse forms is noted $\Omega_{{\rm st}_{f,r}}(\setR^{n+2})$.
\end{definition}
We have proved that, for a given $r$, $m_f{}^*:\ \Omega_{{\rm st}_{f,r}}(\setR^{n+2})\to\Omega(X_f)$ is a bijection, we will see below that $m_f{}^*$, on this space, has a good behavior with respect to usual differential operators.

Note that this definition is equivalent to $\iD\alpha = \iEf\alpha = 0$ and $\LD\alpha =r\alpha$ and $\L_{\sss E_f}\alpha = 0$.

\section{Differential operators on ambient and intrinsic $p$-forms}
\label{sec:Differential}

The correspondence between differential forms on $\setR^{n+2}$ and on $X_f$ finds its realization in the formulas relating the differential calculus on ambient space and on $X_f$.

\subsection{Pull-back of differential operators on $p$-forms}

\begin{property}\label{prop:Restriction}
    Let $\alpha\in\Omega^a(\setR^{n+2})$ be a form of degree $a$, then the pull-backs of (differential) operators acting on $\alpha$ are given by
    \begin{align}
        &m_f{}^*(*_\eta\alpha)
            = *_f\, m_f{}^*(\iF\iD\alpha),         \label{eq:Restr_*}\\
        &m_f{}^*(d\alpha)
            = d_fm_f{}^*\alpha = d_f\alpha_f,    \label{eq:Restr_d}\\
        &m_f{}^*(\delta_\eta\alpha)
            = \delta_f\alpha_f 
                -m_f{}^*[(\LD -2a +n+1)\iF\alpha] \nonumber\\ 
        &\qquad\qquad\qquad\qquad
                -m_f{}^*[(\S^{df} + \square_\eta f - \Fd(\LD -2a +n +1))\iD\alpha], \label{eq:Restr_delta}\\
        &m_f{}^*(\square_\eta\alpha)
            = \square_f\alpha_f 
                +m_f{}^*[(\LD -2a +n-1)\LF\alpha + 2d\iF\alpha]   \nonumber\\
                &\qquad\qquad\qquad\qquad
                    + m_f{}^*[(\S^{df} + \square_\eta f - \Fd(\LD -2a +n -1))\LD\alpha
                    -2\Fd d\iD\alpha]\nonumber\\
                &\qquad\qquad\qquad\qquad
                    + m_f{}^*\{([d,\S^{df} + \square_\eta f] - j^{d\Fd}(\LD -2a +n +1))\iD\alpha\}.
                \label{eq:Restr_Box}
    \end{align}
\end{property}

The proof, while rather straightforward, is somewhat lengthy and has been moved to App.~\ref{proof:Restriction}.
On scalars fields ($a=0$) and in the four-dimensional case with Eq.~\eqref{eq:Restr_Box} we recover Eq.~(9) of Ref.~\onlinecite{Zapata:2017gqg} and Eq.~(13) of Ref.~\onlinecite{Huguet:2022rxi} when restricted to (A)dS spaces.
Similarly, we recover Eq.~(12) of Ref.~\onlinecite{Huguet:2022rxi} for the one-forms ($a=1$) when restricted to (A)dS spaces.
The remaining Schouten--Nijenhuis terms $\S^{df}$ (see App.~\ref{subsec:Schouten-Nijenhuis}) can be expressed in terms of $\LF$ and of the Hessian of $f$ if need be, see Eq.~\eqref{eq:Sdphi}.

Remark that on strongly transverse forms $\alpha_{{\rm st}}$ of degree 0 with respect to $X_f$ one has simply $m_f{}^*(\delta_\eta\alpha_{\rm st}) = \delta_f\alpha_f$ and $m_f{}^*(\square_\eta\alpha_{\rm st}) = \square_f\alpha_f$, as they should.

\begin{property}\label{prop:Hessien}
    Let $\phi$ be a function defined in $\setR^{n+2}$, then the restriction of its Hessian to $X_f$ is given by
    \begin{equation}\label{eq:Hessien}
        m_f{}^*(\nabla^{\eta} d\phi)
        = \nabla^fd_f\phi_f 
        + m_f{}^*[D(\phi) (\nabla^\eta df) + E_f(\phi) \eta], %
    \end{equation}
    with $\nabla$ the Levi-Civita connection.
\end{property}

The proof of Prop.~\ref{prop:Hessien} can be found in App.~\ref{proof:Hessien}.

\begin{remark}
    By setting $\phi = f$ in Eq.~\eqref{eq:Hessien} it reduces to the tautological equation
    $m_f{}^*(\nabla^\eta df) = m_f{}^*(\nabla^\eta df)$,
    since $d_ff_f = d_f 1= 0$, $D(f) = f$ and $F(f) = \Fd$,
    such that $E_f(f) = 0$.
    This makes the Hessian $\nabla^\eta df$ an unavoidable data in the restriction process, see the Riemann tensor and its contractions hereafter.
\end{remark}

\subsection{Extension of differential operators on $p$-forms}

We will now show that operators on spacetime can be computed directly from their extension to ambient space.
The previous section allowed us to calculate the transverse part of $\square_\eta$ on $X_f$.
If we can also calculate the longitudinal part, then it becomes possible, using the decomposition in Eq.~\eqref{Eq:trans-long}, to express $\square_f$ in terms of $\square_\eta$.
That is, to solve the extension problem. This will be possible in the case where the extension is strongly transverse.
For a given $\beta\in\Omega(X_f)$ and $\hat\beta$ its strongly transverse extension of degree zero, we then have the following.

\begin{property}\label{prop:Extension}
    Let $\beta\in\Omega^b(X_f)$ and $\hat\beta$ be its strongly transverse extension of degree zero, we have, on $X_f$
    \begin{align}
    & \delta_f\beta =\delta_\eta\hat\beta,
        \label{eq:delta_fbeta}\\
    & \square_f\beta=
        \square_\eta\hat\beta 
        +m_f{}^l(d\delta_\eta\hat\beta)
        = \square_\eta\hat\beta 
        +(j^{dc}\iEf+j^{df}\iD)d\delta_\eta\hat\beta.
        \label{eq:square_fbeta}
    \end{align}
\end{property}

The proof is given in App.~\ref{sec:Extension}.

\begin{remark}
    Observe that this property can be extended to strongly transverse fields of any degree $r$.
    Nevertheless, this introduces extra terms to Eqs.~\eqref{eq:delta_fbeta} and \eqref{eq:square_fbeta}, leading to a cumbersome result.
\end{remark}

\section{The curvature tensor from $\setR^{n+2}$ and the Weitzenböck formula}
\label{sec:Curvature}

\subsection{The curvature tensor and its contractions}
From the defining function $f$ solely we obtain the Riemann tensor and its contractions.

\begin{property}\label{prop:Riemann_f}
    Using the embedding of $X_f\subset\setR^{n+2}$ described in Sec.~\ref{sec:Geometric_Setting} the Riemann tensor, Ricci tensor, and scalar curvature of $X_f$ are given by
    \begin{align}
        &\Riemann_f 
            = m_f{}^*\bigl\{
                \eta\owedge[(\nabla^\eta df)-\tfrac{1}{2}\Fd\eta]
                \bigr\},\label{eq:Riemann_f}\\
        &\Ricci_f
            = m_f{}^*\bigl[
            (\square_\eta f)\eta
            + (n-2)(\nabla^\eta df)
            -(n-1)\Fd \eta\bigr],\label{eq:Ricci_f}\\
        &\R_f
            = m_f{}^*[2(n-1)(\square_\eta f) -n(n-1)\Fd],\label{eq:R_f}
    \end{align}
    in which the symbol $\owedge$ corresponds to the Kulkarni--Nomizu product [see Eq.~\eqref{eq:Def_KN} and the discussion surrounding it].
\end{property}

The proof of Prop.~\ref{prop:Riemann_f} can be found in App.~\ref{proof:Riemann_f}.
Equations~\eqref{eq:Riemann_f}--\eqref{eq:R_f} provide explicit formulas in ambient space for the Riemann tensor and its contractions.
Equation~\eqref{eq:R_f} for $n=4$ has been found accidentally in an earlier article [Eq.~(10) of Ref.~\onlinecite{Zapata:2017gqg}].

While the ambient space expressions have their own advantages one can reduce them further as is obtained in the following property.

\begin{property}
    Applying $m_f{}^*$ in the formula of the curvature tensor one has, at any point of $X_f$, that
    \begin{align}
        &\Riemann_f 
            = \eta_f\owedge(N_f-\tfrac{1}{2}\Fd\eta_f),
        \label{eq:Riemmann_ff}\\
        &\Ricci_f
            = (\square_\eta f)\eta_f
            + (n-2)N_f
            -(n-1)\Fd \eta_f,\\
        &\R_f
            = 2(n-1)(\square_\eta f) - n(n-1)\Fd,
    \end{align}
    with $\eta_f$ the induced metric on $X_f$ and $N_f = m_f{}^*(\nabla^\eta df)$ such that $N_f(u,v) = -\langle df, \nabla_u^\eta v\rangle$, for $u,v\in T(X_f)$.
\end{property}

\subsection{The Weitzenböck formula}

The formula \eqref{eq:Restr_Box} deals with the Laplace--de~Rham operator.
In order to obtain a similar result concerning the Laplace--Beltrami operator, we must evaluate $\square_f-\Delta_f$, knowing that $\square_\eta-\Delta_\eta=0$.
\begin{property}\label{prop: Weitzenbock}
   For any $\alpha\in\Omega^a(X_f)$, we have
     \begin{align}
        (\square_f-\Delta_f)\alpha
        &=[a(n-a)\Fd + (2a-n)\,{}^{\sss D}N_f-a(\square_\eta f)]\alpha
        \label{eq: LB1}\\
        &=m_f{}^*\{[a(n-a)\Fd + (2a-n)(\S^{df}-\LF)-a(\square_\eta f)]\alpha\}
        \label{eq: LB2}
        \end{align}
    where ${}^{\sss D}N_f=\nabla_\mu e_\nu(f)j^\mu i^\nu$ is the unique derivation on $\Omega(X_f)$ of order 0, vanishing on the scalar fields and coinciding with the contraction with $N_f$ when applied on a 1-form.
\end{property}
The proof is based on the following lemma.
\begin{lemma}\label{lemma:1}
    Let $(M,g)$ be a $n$-dimensional pseudo-Riemaniann manifold such that its Riemann tensor reads:
    \begin{equation}
        \label{eq:Riemann_gT}
        \Riemann= g\owedge T,
    \end{equation}
    where $T$ is a symmetric 2-form.
        Then the Weitzenböck operator is given, for any $\alpha\in\Omega^a(M)$, by
     \begin{equation*}
        (\square-\Delta)\alpha=[(2a-n)\,{}^{\sss D}T-a\Tr(T)]\alpha,
    \end{equation*}    
    where ${}^{\sss D}T = T_{bc}j^bi^c$ is the unique derivation on $\Omega(M)$ of order 0, vanishing on the scalar fields and coinciding with the contraction with $T$ when applied on a 1-form.
\end{lemma}
\begin{remark}
    As is shown in Eq.~\eqref{eq:Riemmann_ff} this applies to all conformally flat spaces since their Riemann tensor is of the form given in Eq.~\eqref{eq:Riemann_gT}.
\end{remark}
The proofs of lemma~\ref{lemma:1} and of property~\ref{prop: Weitzenbock} are available in App.~\ref{proof:Weitzenbock}.

\section{Conformal field equations from $\setR^{6}$ and two-point functions thereof}
\label{sec:Two-point_functions}

As we saw in Sec.~\ref{sec:Differential}, if the field is strongly transverse, the field equations in spacetime correspond to the analogous equations in ambient space, without additional terms. However, for certain equations, the strong transversality condition becomes unnecessary; this is the case for conformal fields, which we will now study.

By utilizing our formalism and previous results, we derive closed-form expressions for the two-point functions of the massless conformally coupled scalar field, of Maxwell's field one-form potential $\alpha$, and of the field strength $\calF$ related to Maxwell's field in any kind of FLRW spacetime.
In the following we reduce the dimensionality of the spacetime to $n=4$, since this is the physically significant scenario and also the case in which Maxwell's equations are conformally invariant.

\subsection{The massless conformally coupled scalar field}

To begin in the simplest case conceivable, let us consider the equation of a massless scalar field coupled to the scalar curvature, using Eqs.~\eqref{eq:Restr_Box} and \eqref{eq:R_f}, we obtain
\begin{multline*}
    (\square_f
    + \gamma \R_f)\phi_f
        = m_f{}^*(\square_\eta\phi)  \\
        - m_f{}^*[2\LF(\LD +1)\phi 
            + (\square_\eta f) (\LD - 6\gamma)\phi
            + \Fd(12\gamma - (\LD+3)\LD)\phi].
\end{multline*}
Now, note that for $\LD\phi = -\phi$ and $\gamma = -\frac{1}{6}$ all the additional terms vanish leaving
\begin{equation*}
    \left(\square_f
        -\tfrac{1}{6} \R_f\right)\phi_f
        = m_f{}^*(\square_\eta\phi).
\end{equation*}
Then, the field $\phi$ can be defined on the ambient space with no further constraint to be fulfilled, such as strong transversality ($\LF\phi=0$) [see the remark following Eq.~\eqref{eq:Restr_Box}].
This is the mark in $\setR^6$ of conformal invariance of fields, being defined on the null cone $\calC$ without imposing any particular conditions involving the defining function $f$.

From earlier work\cite{Huguet:2006fe} we know that the two-point function for this field can be constructed via a mode-sum expansion within a base space $X_f$, for example in the Einstein spacetime $E_{+1}$, as
\begin{equation}
    \label{eq:2ptFunct_Scal}
    \langle\phi(x)\phi(x')\rangle_f
    = \left.\frac{1}{8\pi^2}\left(\frac{1}{y\cdot y'}\right)\right|_{X_f},
\end{equation}
with $y=m_f(x)$, $y'=m_f(x')$ and $y\cdot y'$ the ambient product between two points in $\setR^6$.
We denote the two points in $X_f$ as $x$ and $x'$ in the two-point function to maintain consistency with standard notations.
Additionally, it is to be understood that some $+i\epsilon$ must be added to regularize the function.
Then, exploiting the homogeneity of the field (of degree $-1$), this two-point function can be rescaled from $X_f$ to the null cone $\calC$.  
In this way, one recognizes that the ambient expression~\eqref{eq:2ptFunct_Scal} of the two-point function is valid for arbitrary conformally flat spacetimes $X_f$.

For Minkowski and de Sitter spaces as submanifolds of $\setR^6$, respectively, we have
\begin{align*}
    &y\cdot y'_{|X_0}
    = -\tfrac{1}{2}(\xi_\mu - \xi'_\mu)(\xi^\mu - \xi'^\mu)
    = -\tfrac{1}{2}(\xi - \xi')^2
    = -\tfrac{1}{2}(\Delta\xi)^2,\\
    &y\cdot y'_{|X_{\sss{H^2}}}
    = -H^{-2}(\calZ -1),
\end{align*}
with $\Delta\xi^\mu = \xi^\mu - \xi'^\mu$ and $(\Delta\xi)^2 = \Delta\xi^\mu\Delta\xi_\mu$,
where $\calZ = \cosh(H\mu)$ with $\mu$ the geodesic distance between the two points on de Sitter space (when defined, otherwise see Refs.~\onlinecite{Allen:1985ux,Allen:1985wd} for details and also Ref.~\onlinecite{Huguet:2006fe}).
The interesting cases are, of course, FLRW spaces.
For FLRW spaces of type $k=0$, using Eq.~\eqref{eq:SimpleEmbk=0}, we have
\begin{equation}
    \label{eq:YY_k0}
    y\cdot y'_{|X_{f_{k=0}}}
    = -\tfrac{1}{2}a(\xi^0)a(\xi'^0)(\xi - \xi')^2
    = -\tfrac{1}{2}a(\xi^0)a(\xi'^0)(\Delta\xi)^2.
\end{equation}
Regarding the $k=\pm 1$ cases of the FLRW metric [see Eq.~\eqref{eq:SimpleEmb}], we first reintroduce
\begin{equation*}
    r = \begin{cases}
        \sinh\chi & \text{for $k=-1$,}\\
        \sin\chi  & \text{for $k=+1$,}
    \end{cases}
\end{equation*}
and set $r^i = r\omega^i$, shortened in $\vec{r} = r\vec{\omega}$, such that
\begin{equation}
    \label{eq:YY_k1}
    y\cdot y'_{|X_{f_{k=\pm 1}}}
    = \begin{cases}
        a(t)a(t')[\sqrt{(1+r^2)(1+r'^2)} 
            - \vec{r}\cdot\vec{r'}
            - \cosh(t-t')
        ]
        &\text{for $k=-1$,}\\
        a(t)a(t')[\cos(t-t')
            -\sqrt{(1-r^2)(1-r'^2)} 
            -\vec{r}\cdot\vec{r'}]
        &\text{for $k=+1$.}
    \end{cases}
\end{equation} 
Then, with Eqs.~\eqref{eq:YY_k0} and \eqref{eq:YY_k1} used in Eq.~\eqref{eq:2ptFunct_Scal} we obtain the two-point function of the massless conformally coupled field in all types of FLRW spaces.
Of course many expressions of this two-point function are known,\cite{Birrell:1982ix}  the appeal of the expression presented here lies in its simplicity, which is a direct product of the simplicity of the embedding formulas of FLRW spaces in $\setR^6$ [Eq.~\eqref{eq:SimpleEmb}].
This is particularly useful in avoiding the use of awkward scale factors, like those presented in Eqs.~\eqref{eq:Akpm1} and \eqref{eq:A_Exponential}.

\begin{remark}
    From its ambient expression in Eq.~\eqref{eq:2ptFunct_Scal} the scalar two-point function is manifestly $SO(2,4)$ invariant.
    While this invariance remains once brought on $X_f$ we lose its \emph{manifest} invariance.
    However, from Eq.~\eqref{eq:YY_k0}, we easily read off the $E(3)$ invariance of the $k=0$ case of FLRW spaces.
    Rewriting Eq.~\eqref{eq:YY_k1} as
    \begin{equation*}
        y\cdot y'_{|X_{f_{k=\pm 1}}}
        = \begin{cases}
            a(t)a(t')[\cosh\chi\cosh\chi'
                - \sinh\chi\sinh\chi' \vec{\omega}\cdot\vec{\omega'}
                - \cosh(t-t')
            ]
            &\text{for $k=-1$,}\\
            a(t)a(t')[\cos(t-t')
                -(\cos\chi\cos\chi'
                + \sin\chi\sin\chi'\vec{\omega}\cdot\vec{\omega'})]
            &\text{for $k=+1$,}
        \end{cases}
    \end{equation*}
    we recognize in the $k=-1$ case the $SO(1,3)$ invariant product on the hyperboloid $\mathbb{H}^{3}$.
    Identically, in the $k=+1$ case we recognize the $SO(4)$ invariant scalar product on the sphere $\mathbb{S}^3$.
    See App.~\ref{sec:IsoFLRW} for a discussion of additional isometries of FLRW spaces from our  ambient space point of view.
\end{remark}

\subsection{Maxwell's field}

\subsubsection{The potential $\alpha$}

Maxwell's equations are the proverbial example\cite{Cunningham:1910pxu,Bateman:1909pyp} of conformal invariance in physics.
In our ambient formalism in $\setR^6$ the equation on the one-form potential $\alpha$ on $X_f$, using Eqs.~\eqref{eq:Restr_d} and \eqref{eq:Restr_delta}, reads as 
\begin{align*}
    \delta_f d_f\alpha_f
    &= m_f{}^*(\delta_\eta d_\eta\alpha)
        +m_f{}^*[(\LD +1)\iF d \alpha 
        +(\S^{df} + \square_\eta f - \Fd(\LD +1))\iD d \alpha]\\
    &= m_f{}^*(\delta_\eta d_\eta\alpha)
        +m_f{}^*[\iF d\LD \alpha 
        +(\S^{df} + \square_\eta f - \Fd(\LD +1))(\LD - d\iD)\alpha].
\end{align*}
Thus, for $\LD\alpha = 0$ and $\iD\alpha = 0$  it reduces to
\begin{equation}
    \label{eq:Maxwell_A}
    \delta_f d_f\alpha_f
    = m_f{}^*(\delta_\eta d_\eta\alpha),
\end{equation}
with again the fact that $\alpha$ does not need to be transverse.
Note that this is true only for $n=4$, Maxwell's  equations are not conformally invariant outside of this case.

However, Maxwell's field differs from the massless conformally coupled scalar field in that it is gauge invariant.
Each solution $\alpha$ to Maxwell's equations is determined up to a gauge term $d\phi$, with $\phi$ an arbitrary scalar field.
This is an obstruction to obtaining straightforwardly two-point functions.
To circumvent this hurdle one needs to invoke a quantization process dealing with this gauge invariance, popular choices are Gupta\cite{Gupta:1949rh}-Bleuler\cite{Bleuler:1950cy} or Becchi-Rouet-Stora-Tyutin (see Refs.~\onlinecite{Becchi:1996yh,Zinn-Justin:1999euj} and references therein) schemes.
However, in nearly all works concerned with Maxwell's field and seeking a propagator for the photon, conformal invariance is broken.
This needs not to be the case.
First, there is a conformally invariant gauge fixing equation: the Eastwood-Singer gauge.\cite{Eastwood:1985eh}
Then, exploiting the fields in $\setR^6$, there is a modification of Nakanishi\cite{Nakanishi:1966zz} kind of the Gupta-Bleuler scheme leading to a conformally covariant quantization of the potential $\alpha$.
Without the details leading to it (which are out of the scope of the current article, see Refs.~\onlinecite{Binegar1982AdS,Binegar1982Mink,Bayen:1984dt,Faci2009,queva:tel-00503186}) the two-point function reads as 
\begin{equation}
    \label{eq:<aa>}
    \langle \alpha(x)\alpha(x')\rangle_f
    = -\frac{1}{8\pi^2}m_f{}^*\left(
    \frac{\eta_{\alpha\beta'} dy^\alpha \otimes dy'^{\beta'}}
        {y\cdot y'}\right),
\end{equation}
see commentaries after Eq.~\eqref{eq:2ptFunct_Scal} about the notations.

From its expression in ambient space and our embedding formulas we can expand its applicability from Minkowski and de Sitter spaces to FLRW spaces.
In order to compute $\eta_{\alpha\beta'} dy^\alpha\otimes dy'^{\beta'}$ it is simpler to set $y^\alpha = a(t)y_{\sss{E_k}}^\alpha$ with $y_{\sss{E_k}}$ coordinates on the Einstein space $E_k$ corresponding to the $k$ type of the FLRW space (which for $k=0$ is Minkowski space).
We then have
\begin{multline}
    \label{eq:dy.dy'}
    \eta_{\alpha\beta'}dy^\alpha\otimes dy'^{\beta'}
    = a(t)a(t')[\eta_{\alpha\beta'} dy_{\sss{E_k}}^\alpha\otimes dy'_{\sss{E_k}}{}^{\beta'}
        + \dot{\psi}(t)dt\otimes y_{\sss{E_k}\beta'} dy'_{\sss{E_k}}{}^{\beta'}\\
        + y'_{\sss{E_k}\alpha} dy_{\sss{E_k}}^\alpha \otimes \dot{\psi}(t') dt
        + y_{\sss{E_k}}\cdot y'_{\sss{E_k}}
            \dot{\psi}(t)\dot{\psi}(t') dt\otimes dt'],
\end{multline}
where we set $\dot{\psi} = \frac{\dot{a}}{a} = \frac{d\ }{dt} \ln a = \frac{d\ }{dt} \psi$ the logarithmic derivative of the conformal scale factor.

Regardless of the chosen FLRW space type and without specific computations, we can draw very strong conclusions about the two-point function $\langle\alpha(x)\alpha(x')\rangle_{\sss{FLRW}_k}$.
From Eqs.~\eqref{eq:<aa>} and \eqref{eq:dy.dy'}, taking into account the cancellation of the scale factors $a(t)a(t')$ by the denominator, the two-point function on an FLRW space of type $k$ is that of the associated Einstein space $E_k$, albeit with additional terms.
Concerning these additional terms, note that the last additional term can be written as
\begin{equation*}
    -\frac{1}{8\pi^2}\dot{\psi}(t)\dot{\psi}(t')dt\otimes dt'
    = -\frac{1}{8\pi^2} d\psi(t)\otimes d\psi(t')
    = -\frac{1}{8\pi^2} dd' [\psi(t)\psi(t')],
\end{equation*}
which is manifestly a pure gauge term at $y$ and $y'$.
Similarly, we note that
\begin{align*}
    \frac{a(t)a(t')}{y\cdot y'} \dot{\psi}(t)dt\otimes y_{\sss{E_k}}\cdot dy'_{\sss{E_k}}
    &= \frac{1}{y_{\sss{E_k}}\cdot y'_{\sss{E_k}}}
        d\psi(t)\otimes y_{\sss{E_k}}\cdot dy'_{\sss{E_k}}\\
    &= d\psi(t)\otimes d' \ln(y_{\sss{E_k}}\cdot y'_{\sss{E_k}})
    = d'[\ln(y_{\sss{E_k}}\cdot y'_{\sss{E_k}}) d\psi(t)],
\end{align*}
which is manifestly a pure gauge term at $y'$.
The same symmetrically holds for the remaining additional term, which is a pure gauge term at $y$.
In summary, all additional terms to the two-point function on the underlying Einstein space are pure gauge terms and one can thus choose to discard them.
Therefore, we can write the two-point function as
\begin{equation*}
    \langle\alpha(x)\alpha(x')\rangle_{\sss{FLRW}_{k}}
    = \langle\alpha(x)\alpha(x')\rangle_{\sss{E}_{k}}
    + PG_k(x,x')
\end{equation*}
and in practical computations neglect the pure gauge term $PG_k(x,x')$.
The ambient space formulation clearly reveals that every extra contribution to the two-point function is a pure gauge term, a feature that is difficult to track when one works only with intrinsic coordinates on the chosen FLRW background.

We stated this property on FLRW spaces, but it holds for all conformally flat spaces in which the two-point function on one space is equal to that on a chosen base space plus additional pure gauge terms arising from the rescaling in $\setR^6$ from this base space.
Indeed, it suffices to consider $a$ as a function not only of $t$ and set similarly $da = a (da/a) = a d\ln a = a d\psi$.

Now, it is sufficient to calculate the two-point function for each underlying Einstein space in order to express it in the coordinates most suitable for practical computations on FLRW spaces.
To avoid confusion, let us recall that in this article $t$ denotes the \emph{conformal time}.
For $k=0$ the corresponding Einstein space is Minkowski space and the two-point function is
\begin{align}
    \langle\alpha(x)\alpha(x')\rangle_{\sss{E_0}}
    &= -\frac{1}{8\pi^2}\frac{1}{(y_{\sss{E_0}}\cdot y'_{\sss{E_0}})}
        (dt\otimes dt' - \sum_i dr^i\otimes dr'^i)\nonumber\\
    &= \frac{1}{4\pi^2}\frac{1}{(\Delta\xi)^2} d\xi^\mu \otimes d\xi'_\mu,
    \label{eq:<aa>_k0}
\end{align}
which is a well-known result and coincides with the one usually obtained in Lorenz gauge.
For $k=-1$ the two-point function reads as
\begin{multline}
    \label{eq:<aa>_k-1}
    \langle\alpha(x)\alpha(x')\rangle_{\sss{E_{-1}}}
    = -\frac{1}{8\pi^2}\frac{1}{(y_{\sss{E}_{-1}}\cdot y'_{\sss{E}_{-1}})}\Bigl[
        \cosh(t-t')dt\otimes dt'
        - \sum_i dr^i\otimes dr'^i\\
        + \frac{1}{\sqrt{(1+r^2)(1+r'^2)}}\sum_{ij} r^i dr^i\otimes r'^j dr'^j
    \Bigr],
\end{multline}
with $(y_{\sss{E}_{-1}}\cdot y'_{\sss{E}_{-1}})$ as in Eq.~\eqref{eq:YY_k1} with $a=1$.
Finally, for $k=+1$ it reads as
\begin{multline}
    \label{eq:<aa>_k+1}
    \langle\alpha(x)\alpha(x')\rangle_{\sss{E_{+1}}}
    = -\frac{1}{8\pi^2}\frac{1}{(y_{\sss{E}_{+1}}\cdot y'_{\sss{E}_{+1}})}\Bigl[
        \cos(t-t')dt\otimes dt'
        - \sum_i dr^i\otimes dr'^i\\
        - \frac{1}{\sqrt{(1-r^2)(1-r'^2)}}\sum_{ij} r^i dr^i\otimes r'^j dr'^j
    \Bigr],
\end{multline}
with $(y_{\sss{E}_{+1}}\cdot y'_{\sss{E}_{+1}})$ as in Eq.~\eqref{eq:YY_k1} with $a=1$.
Or one can adopt as the two-point function for the potential $\alpha$ the one on de Sitter space\cite{Faci2009,queva:tel-00503186} which, in coordinates, reads as
\begin{equation}
    \label{eq:<aa>dS}
    \langle\alpha_{\mu}(x)\alpha_{\nu'}(x')\rangle_{\sss{\text{dS}}}
    = \frac{H^2}{8\pi^2}\left(
        \frac{1}{\calZ-1}g_{\mu\nu'} -n_\mu n_{\nu'}\right),
\end{equation}
with $g_{\mu\nu'}$ the parallel propagator along the geodesic linking points $p$ and $p'$, $n_\mu$ and $n_{\nu'}$ the unit tangent vectors to the geodesic at points $p$ and $p'$ (see Ref.~\onlinecite{Allen:1985wd} for details),
and then adopt coordinates putting the dS metric into a FLRW metric of chosen $k$ type [see Eqs.~\eqref{eq:dS_k=-1}--\eqref{eq:dS_k=1} in App.~\ref{subsec:Improved_dS}].
From this perspective, since all these two-point functions are all related to each other up to pure gauge terms, selecting a specific two-point function ultimately becomes a matter of expedience and convenience with respect to practical computations.
That is, depending on the coordinates used, one can select the most useful expression between Eqs.~\eqref{eq:<aa>_k0} and \eqref{eq:<aa>dS}.

\begin{remark}
    The added pure gauge terms could be useful if one wants to consider a flat space limit (for $k=\pm 1$) or keep the $SO(2,4)$ invariance of the two-point function (which is broken by removing those terms).
    To those terms one could even add other pure gauge terms, e.g., to improve the large-distance behavior, etc.
    Thus, for completeness, we record below the three pure gauge terms $PG_k$, which can be added to the two-point functions given in Eqs.~\eqref{eq:<aa>_k0}--\eqref{eq:<aa>_k+1}.
    \begin{align*}
        &-8\pi^2PG_0(x,x') = d\psi(t)\otimes d\psi(t')
        + \frac{2}{(\Delta\xi)^2}(\Delta\xi_\mu)\Bigl[
            d\xi^\mu\otimes d\psi(t')
            - d\psi(t)\otimes d\xi'^\mu\Bigr],\\
        &-8\pi^2PG_{-1}(x,x') = d\psi(t)\otimes d\psi(t')
            + \frac{1}{(y_{\sss{E}_{-1}}\cdot y'_{\sss{E}_{-1}})}\Bigl[\\
        &\qquad\qquad\qquad\qquad\qquad
            d\psi(t)\otimes\Bigl(\sinh(t-t')dt'
                    -\sum_i \Bigl(r^i - r'^i\sqrt{\frac{1+r^2}{1+r'^2}} \Bigr)dr'^i\Bigr)\\
        &\qquad\qquad\qquad\qquad\qquad
            -\Bigl(\sinh(t-t')dt
                    +\sum_i \Bigl(r'^i - r^i\sqrt{\frac{1+r'^2}{1+r^2}} \Bigr)dr^i\Bigr)
            \otimes d\psi(t')
            \Bigr],\\
        &-8\pi^2PG_{+1}(x,x') = d\psi(t)\otimes d\psi(t')
            + \frac{1}{(y_{\sss{E}_{+1}}\cdot y'_{\sss{E}_{+1}})}\Bigl[\\
        &\qquad\qquad\qquad\qquad\qquad
            d\psi(t)\otimes\Bigl(\sin(t-t')dt'
                    -\sum_i \Bigl(r^i - r'^i\sqrt{\frac{1-r^2}{1-r'^2}} \Bigr)dr'^i\Bigr)\\
        &\qquad\qquad\qquad\qquad\qquad
            -\Bigl(\sin(t-t')dt
                    +\sum_i \Bigl(r'^i - r^i\sqrt{\frac{1-r'^2}{1-r^2}} \Bigr)dr^i\Bigr)
            \otimes d\psi(t')
            \Bigr].
    \end{align*}
\end{remark}

\subsubsection{The field strength $\calF$}

The field strength $\calF$ being a $2$-form we obtain, from Eqs.~\eqref{eq:Restr_d} and \eqref{eq:Restr_delta}, that Maxwell's equations for $\calF$ read as
\begin{align*}
    &\delta_f\calF_f
    = m_f{}^*(\delta_\eta\calF )
        +m_f{}^*[\iF\LD\calF
        +(\S^{df} + \square_\eta f - \Fd(\LD + 1))\iD\calF],\\
    &d_f\calF_f = m_f{}^*(d\calF),    
\end{align*}
and extend to the null cone $\calC$ with no particular constraints with respect to $f$ to be fulfilled if $\LD\calF = 0$ and $\iD\calF = 0$.
This is, of course, compatible with the conformal invariance of Maxwell's equations for the potential [Eq.~\eqref{eq:Maxwell_A}] with $\calF = d\alpha$.

Using the fact that $d_fm_f{}^* = m_f{}^*d_\eta$ [Eq.~\eqref{eq:Restr_d}] the two-point function on the field strength can easily be obtained from the ambient space, it is given by
\begin{align}
    \langle \calF(x)\calF(x')\rangle_f
    &= d_f d'_f\langle \alpha (x)\alpha(x')\rangle_f
    = \langle d_\eta\alpha (x) d'_\eta\alpha(x')\rangle_f\nonumber\\
    &= \frac{1}{8\pi^2}m_f{}^*\biggl(
    \frac{\eta_{\alpha\beta'}\eta_{\gamma \delta'}}{(y\cdot y')^2}
        - \frac{2}{(y\cdot y')^3}
            \eta_{\alpha\beta'} y'_\gamma y_{\delta'}
    \biggr) dy^\alpha\wedge dy^\gamma\otimes dy'^{\beta'}\wedge dy'^{\delta'},
    \label{eq:<FF>}
\end{align}
in which we used the shorthands $y'_\alpha = \eta_{\alpha\beta'} y'^{\beta'}$ and $y_{\beta'} = \eta_{\alpha\beta'}y^\alpha$.

\begin{remark}
In the parenthesis of the two-point function given in Eq.~\eqref{eq:<FF>} we can anti-symmetrize explicitly the components.
Then, by recognizing their algebraic symmetries, we can rewrite entirely the two-point function for the field strength by using \emph{formally} the Kulkarni--Nomizu product,
\begin{align*}
    \langle \calF(x)\calF(x')\rangle_f
    &= \frac{1}{16\pi^2}m_f{}^*\biggl(
        \frac{1}{(y\cdot y')^2}
            (\eta_{\alpha\beta'}\eta_{\gamma \delta'}
                - \eta_{\gamma\beta'}\eta_{\alpha\delta'} )\\
    &\qquad\qquad
            - \frac{1}{(y\cdot y')^3}
                (\eta_{\alpha\beta'} y'_\gamma y_{\delta'}
                -\eta_{\gamma\beta'} y'_\alpha y_{\delta'}
                -\eta_{\alpha\delta'} y'_\gamma y_{\beta'}
                +\eta_{\gamma\delta'} y'_\alpha y_{\beta'})\biggr)\times\\
    &\qquad\qquad\times dy^\alpha\wedge dy^\gamma\otimes dy'^{\beta'}\wedge dy'^{\delta'}\\
    &= \frac{1}{16\pi^2}m_f{}^*\biggl(
        \frac{1}{(y\cdot y')^2}\frac{1}{2} 
            (\eta_{\alpha\beta'}dy^\alpha\otimes dy'^{\beta'})\owedge
            (\eta_{\gamma\delta'}dy^\gamma\otimes dy'^{\delta'})\\
    &\qquad\qquad
            - \frac{1}{(y\cdot y')^3}
            (\eta_{\alpha\beta'}dy^\alpha\otimes dy'^{\beta'})\owedge
            (y'_\gamma dy^\gamma\otimes y_{\delta'} dy'^{\delta'})
        \biggl).%
\end{align*} 
\end{remark}

Note that, for $X_{f_2}$ and $X_{f_1}$ two spacetimes with $X_{f_1}$ seen as a base space, since one locally has
    $\langle\alpha(x)\alpha(x')\rangle_{f_2} 
        = \langle\alpha(x)\alpha(x')\rangle_{f_1} + PG(x,x')$,
applying $d$ and $d'$ cancels the pure gauge terms and one obtains the invariance of the two-point function for the field strength,
\begin{equation*}
    \langle \calF(x)\calF(x')\rangle_{f_2}
    =  \langle \calF(x)\calF(x')\rangle_{f_1}.
\end{equation*}
This is to be expected, the two-point functions for the field strength across all conformally flat spaces are identical.
This is an incarnation of the conformal invariance of zero weight for this two-point function, allowing it to be rescaled from Minkowski space, or from any conformally flat space chosen as a base space.
Here this confirms that our two-point functions $\langle\alpha(x)\alpha(x')\smash{\rangle_{\sss{FLRW}_k}}$ for the potential on all FLRW spaces [Eqs.~\eqref{eq:<aa>_k0}--\eqref{eq:<aa>dS}] carry in themselves the physical content leading to the expected  two-point function for the field strength.

Then, it suffices to compute the two-point function on Minkowski space, which is calculated below and given in Eq.~\eqref{eq:<FF>0}.
However, we believe that the explicit expression for the $k=-1$ type of FLRW space [Eqs.~\eqref{eq:FF_{0i,0m}_k-1}--\eqref{eq:FF_{ij,mn}_k-1}] and for the $k=+1$ type [Eqs.~\eqref{eq:FF_{0i,0m}_k+1}--\eqref{eq:FF_{ij,mn}_k+1}] in the usual coordinates on these spaces can also be useful.

For $k=0$ the two-point function is best expressed in Cartesian coordinates $\xi$ and computed directly on Minkowski space.
Regarding the computation we note that
\begin{align*}
    dy_{0\,\alpha}\wedge dy_{0\,\beta}\otimes
        dy'_0{}^\alpha\wedge dy'_0{}^\beta
    &= - 2 \sum_i dt\wedge dr^i \otimes dt'\wedge dr'^i
    + \sum_{ij} dr^i \wedge dr^j \otimes dr'^i\wedge dr'^j\\
    &= d\xi^\mu\wedge d\xi^\nu \otimes d\xi'_\mu\wedge d\xi'_\nu,
\end{align*}
and
\begin{align*}
    dy_{0\,\alpha} \wedge y'_{0\,\beta} dy_0{}^\beta
        \otimes dy'_0{}^\alpha &\wedge y_{0\,\gamma} dy'_0{}^\gamma
    = \sum_i (\Delta t)^2 dt\wedge dr^i\otimes dt'\wedge dr'^i\\
    &- \sum_{ij} (\Delta r^i) (\Delta r^j)
        dt\wedge dr^i\otimes dt'\wedge dr'^j\\
    &- \sum_{ij} (\Delta t)(\Delta r^i) [dt\wedge dr^j\otimes dr'^i\wedge dr'^j
        + dr^i\wedge dr^j\otimes dt'\wedge dr'^j ]\\
    &+ \sum_{ijk} (\Delta r^i)(\Delta r^j) dr^i\wedge dr^k
        \otimes dr'^j\wedge dr'^k\\
    &= -(\Delta\xi_\mu)(\Delta\xi_\nu)
        d\xi^\mu\wedge d\xi^\rho\otimes d\xi'^\nu\wedge d\xi'_\rho,
\end{align*}
with $\Delta t = t-t'$ and $\Delta r^i = r^i - r'^i$,
leading to the compact, manifestly Poincaré invariant, two-point function
\begin{multline}
    \label{eq:<FF>0}
    \langle \calF(x)\calF(x')\rangle_{\sss{E_0}}
    = \frac{1}{2\pi^2}\frac{1}{(\Delta\xi)^4}
        d\xi^\mu\wedge d\xi^\nu \otimes d\xi'_\mu\wedge d\xi'_\nu\\
        - \frac{2}{\pi^2}\frac{1}{(\Delta\xi)^6}
            (\Delta\xi_\mu)(\Delta\xi_\nu)
        d\xi^\mu\wedge d\xi^\rho\otimes d\xi'^\nu\wedge d\xi'_\rho,
\end{multline}
writing $(\Delta\xi)^4 = [(\Delta\xi)^2]^2 = [(\Delta\xi_\mu)(\Delta\xi^\mu)]^2$ and similarly $(\Delta\xi)^6 = [(\Delta\xi)^2]^3$.
This is, naturally, the familiar Minkowskian two-point function for the field strength, derived from the equally familiar two-point function for the potential [Eq.~\eqref{eq:<aa>_k0}].

For $k=-1$, unlike the $k=0$ case, the two-point function for the field strength does not admit a compact expression.
We write it out in terms of its components.
First, we define the functions
\begin{align*}
    &D_{im} = \delta_{im} - \frac{r^ir'^m}{\sqrt{(1+r^2)(1+r'^2)}},\\
    &G_{im} = r^i r'^m + r'^i r^m
            -\sqrt{\frac{1+r'^2}{1+r^2}} r^i r^m
            -\sqrt{\frac{1+r^2}{1+r'^2}} r'^i r'^m.
\end{align*}
Then, we obtain
\begin{align}
    &4\pi^2\langle\calF_{0i}(x)\calF_{0m}(x')\rangle_{\sss{E}_{-1}} %
    = -\frac{\cosh(t-t')}{(y\cdot y')^2} D_{im}
        -\frac{1}{(y\cdot y')^3}\biggl[
            \cosh(t-t')G_{im}
             +\sinh^2(t-t') D_{im}
        \biggr],
        \label{eq:FF_{0i,0m}_k-1}\\
    &8\pi^2\langle\calF_{0i}(x)\calF_{mn}(x')\rangle_{\sss{E}_{-1}} %
    = \frac{\sinh(t-t')}{(y\cdot y')^3}\ \biggl[r^m\ D_{in} - r^n D_{im}
    + \sqrt{\frac{1+r^2}{1+r'^2}}(r'^m\delta_{in} - r'^n\delta_{im})
    \biggr],
    \label{eq:FF_{0i,mn}_k-1}\\
    &16\pi^2\langle\calF_{ij}(x)\calF_{mn}(x')\rangle_{\sss{E}_{-1}} %
    = \frac{1}{(y\cdot y')^2}
        (D_{im}\delta_{jn}
        - D_{jm}\delta_{in}
        - D_{in}\delta_{jm}
        + D_{jn}\delta_{im}
        - \delta_{im}\delta_{jn}
        + \delta_{jm}\delta_{in}
        )\nonumber\\ &\qquad\qquad
    - \frac{1}{(y\cdot y')^3}\biggl[
    \frac{(r'^i r^j r^m r'^n
        - r'^j r^i r^m r'^n
        - r'^i r^j r^n r'^m
        + r'^j r^i r^n r'^m)}
    {\sqrt{(1+r^2)(1+r'^2)}}
    \nonumber\\
    &\qquad\qquad\qquad\qquad\qquad
    - (G_{im}\delta_{jn} - G_{jm}\delta_{in} - G_{in}\delta_{jm} + G_{jn}\delta_{im})
    \biggr],
    \label{eq:FF_{ij,mn}_k-1}
\end{align}
with $y\cdot y' = y_{\sss{E_{-1}}}\cdot y'_{\sss{E_{-1}}}$ as in Eq.~\eqref{eq:YY_k1} with $a=1$
and where we carried out explicitly the anti-symmetrization of the indices.
The components $\langle\calF_{ij}(x)\calF_{0m}(x')\smash{\rangle_{\sss{E}_{-1}}}$ of the two-point function are obtained from  $\langle\calF_{0i}(x)\calF_{mn}(x')\smash{\rangle_{\sss{E}_{-1}}}$ by simultaneously switching each set of indices and the primes.

For $k=+1$, this is very similar to the $k=-1$ case, we define 
\begin{align*}
    &D_{im} = \delta_{im} + \frac{r^ir'^m}{\sqrt{(1-r^2)(1-r'^2)}},\\
    &G_{im} = r^i r'^m +  r'^ir^m
        -  \sqrt{\frac{1-r'^2}{1-r^2}}r^ir^m
        -  \sqrt{\frac{1-r^2}{1-r'^2}}r'^ir'^m
\end{align*} 
and the two-point function in components reads as
\begin{align}
    &4\pi^2\langle\calF_{0i}(x)\calF_{0m}(x')\rangle_{\sss{E}_{+1}} %
    = -\frac{\cos(t-t')}{(y\cdot y')^2} D_{im}
    -\frac{1}{(y\cdot y')^3}\biggl[
        \sin^2(t-t')D_{im}
        + \cos(t-t')G_{im}
    \biggr],
    \label{eq:FF_{0i,0m}_k+1}\\
    &8\pi^2\langle\calF_{0i}(x)\calF_{mn}(x')\rangle_{\sss{E}_{+1}} %
    = \frac{\sin(t-t')}{(y\cdot y')^3}\biggl[
        r^m D_{in} - r^n D_{im}
        - \sqrt{\frac{1-r^2}{1-r'^2}} (r'^m\delta_{in} - r'^n\delta_{im})
    \biggr],
    \label{eq:FF_{0i,mn}_k+1}\\
    &16\pi^2\langle\calF_{ij}(x)\calF_{mn}(x')\rangle_{\sss{E}_{+1}} %
    = \frac{1}{(y\cdot y')^2}
    (\delta_{im}D_{jn}
    - \delta_{jm}D
    _{in}
    - \delta_{in}D_{jm}
    + \delta_{jn}D_{im}
    + \delta_{im}\delta_{jn} - \delta_{jm}\delta_{in})
    \nonumber\\ &\qquad\qquad
    +\frac{1}{(y\cdot y')^3}\biggl[
        \frac{(
            r'^ir^jr^mr'^n
            - r'^jr^ir^mr'^n
            - r'^ir^jr^nr'^m
            + r'^jr^ir^nr'^m)}{\sqrt{(1-r^2)(1-r'^2)}}
    \nonumber\\
    &\qquad\qquad\qquad\qquad\qquad
        + G_{im} \delta_{jn}
        - G_{jm} \delta_{in}
        - G_{in} \delta_{jm}
        + G_{jn} \delta_{im}
    \biggr].
    \label{eq:FF_{ij,mn}_k+1}
\end{align}

\begin{remark}
    On de Sitter space one can get from Eq.~\eqref{eq:<aa>dS}  that the two-point function, in components, reads as\cite{Allen:1985wd,Faci2009,queva:tel-00503186}
    \begin{equation}
        \label{eq:<FF>dS}
        \langle\calF^{\mu\nu}(x)\calF_{\rho'\sigma'}(x')\rangle_{\sss{\text{dS}}}
        = \frac{1}{8\pi^2}
        \frac{1}{[H^{-2}(\calZ-1)]^2}\left(
            g^{[\mu}{}_{[\rho'} g^{\nu]}{}_{\sigma']}
            - 4n^{[\mu}g^{\nu]}{}_{[\rho'}n_{\sigma']}
        \right).
    \end{equation}
    Again, as all these two-point functions are the same, the task is to choose the most advantageous form based on the computational context.
    That is, depending on the coordinates used, between Eqs.~\eqref{eq:<FF>0} and \eqref{eq:<FF>dS}.
\end{remark}

\section{Conclusion}
\label{sec:Conclusion}

We investigate in this article two primary aspects of $n$-dimensional locally conformally flat spaces.
First, we develop differential-geometric methods to understand these spaces as submanifolds in the ambient space $\setR^{n+2}$.
Second, this approach is applied to FLRW geometries, leading to simplified expressions for the photon propagator in four dimensions.

The paper highlights the relationship between ambient and intrinsic quantities through differential geometry (Props.~\ref{prop:Restriction}--\ref{prop: Weitzenbock}), emphasizing the need for explicit embedding formulas for various spacetimes.
A central result is the derivation of explicit and strikingly simple embeddings of general FLRW spacetimes of any $k$ type ($k=\pm 1,0$) into $\mathbb{R}^{n+2}$ [Eq.~\eqref{eq:SimpleEmb} and Prop.~\ref{prop:SimpleEmb}].
These embeddings appear significantly simpler than those available in the existing literature. 
Within this setting, we derive concise expressions for two-point functions of conformally invariant fields, including an especially simple ambient-space representation of the photon propagator in four-dimensional FLRW spacetimes [Eq.~\eqref{eq:<aa>}].
This leads to explicit and simplified formulas for the photon propagator on FLRW spaces of any $k$ type [Eqs.~\eqref{eq:<aa>_k0}--\eqref{eq:<aa>_k+1}].
We hope that these embeddings and expressions of two-point functions of conformally invariant fields can help with the investigation of cosmological phenomena.

\appendix
\addcontentsline{toc}{section}{Appendixes}
\section{Useful technical tools and reminder of needed differential geometry formulas}
\label{sec:Useful_formulas}

\subsection{Summary of our notations in differential geometry\label{diff.geom.}}

Here, we bring together our notations and conventions, which mostly follow those of Ref.~\onlinecite{Fecko}. 

Let $(M,g)$ be a $n$-dimensional oriented pseudo-Riemannian manifold with $\nabla^g$ its Levi-Civita connection.
We note $\Omega^p(M) = \smash{\bigwedge^p T^*(M)}$ and $\Omega(M)=\sum_p\Omega^p(M)$ the sets of differential forms on $M$ and $\tilde g$ the corresponding metric on $\Omega(M)$.
The volume form is written $\omega_g$ and the Hodge operator $*_g$ is defined, as usual, through $\alpha_1\wedge *_g\alpha_2=\tilde g(\alpha_1,\alpha_2)\omega_g$ where $\alpha_1,\alpha_2\in\Omega^p(M)$.
The natural pairing between a $1$-form $\lambda$ and a vector $v$ is written $\langle\lambda,v\rangle$, the musical applications $\flat_g$ and $\sharp_g$ relate this pairing to the metric as $\langle\lambda,v\rangle = g(\sharp_g\lambda,v)$ and 
$g(u,v) = \langle\flat_g u,v\rangle.$

The codifferential operator $\delta_g$ is defined by
\begin{equation*}
    \delta_g\alpha=(-1)^a*_g^{-1}d*_g \alpha,
\end{equation*}
for $\alpha\in\Omega^a(M)$.
The de Rham and Beltrami Laplacians are, respectively, defined by
\begin{equation*}
    \square_g=-(d\delta_g+\delta_gd),\qquad
    \Delta_g = \Tr(\nabla^g)^2,
\end{equation*}
where $(\nabla^g)^2$ is the second covariant derivative.
The Hessian $\nabla^gdf$, which is symmetric, is defined by
\begin{equation*}
    (\nabla^g d\phi)(u,v)
    =\langle\nabla^g_u d\phi,v\rangle
    =(uv-\nabla^g_uv)(\phi)
    =(\nabla^g d\phi)(v,u).
\end{equation*}

Of prime importance here are the ``creator operator'' $j_v$, such that $j_v\alpha = (\flat_g v)\wedge\alpha$ with $v\in T(M)$ and $\alpha\in\Omega^a$, and the interior product $i_v$ (or ``annihilation'' operator $i_v$). 
They fulfill
\begin{align*}
    &j_uj_v = -j_vj_u,\qquad
    i_ui_v = -i_vi_u,\qquad
    i_u j_v + j_vi_u = g(u,v),\\
    &i_v\alpha\wedge\beta = (i_v\alpha)\wedge\beta + (-1)^a\alpha\wedge(i_v\beta),\quad
    \alpha\in\Omega^a,\ \beta\in\Omega^b.
\end{align*}
These interior and exterior products $i_v$ and  $j_v$ are extended to $1$-forms in a natural way.
For $\lambda\in\Omega^1$ and $\beta\in\Omega^b$ we note
\begin{align*}
    &i^\lambda\beta = i_{\sharp_g\lambda}\beta,\\
    &j^\lambda\beta = j_{\sharp_g\lambda}\beta = \lambda\wedge\beta.
\end{align*}
They correspond to each other through the Hodge operator
\begin{equation}\label{eq:springboard1}
    i^\lambda\beta = (-1)^{b+1} *_g^{-1}j^\lambda*_g\beta.
\end{equation}

For any basis $e_a$ together with the dual basis $e^b$ we note
\begin{equation*}
    i_a=i_{e_a},\qquad i^b=i^{e^b}=\tilde g^{ba}i_a,\qquad
    j^b=j^{e^b},\qquad j_a=j_{e_a}=g_{ab}j^b.
\end{equation*}

\subsection{Schouten--Nijenhuis bracket on forms}
\label{subsec:Schouten-Nijenhuis}

The Schouten--Nijenhuis bracket can be brought from multivectors to forms using musical operators.\cite{Huguet:2024mza}
While dealing with codifferential of products between differential forms this bracket naturally turns up.
Indeed, for $\alpha\in\Omega^a$ and $\beta\in\Omega^b$ it can be defined as
\begin{equation*}
    \llbra\alpha,\beta\rrket
    = (-1)^a[\delta_g(\alpha\wedge\beta) - (\delta_g\alpha)\wedge\beta -(-1)^a\alpha\wedge(\delta_g\beta)],
\end{equation*}
with $\llbra\alpha,\beta\rrket$ the bracket between $\alpha$ and $\beta$, which will be conveniently denoted as
\begin{equation*}
    \S^\alpha\beta = \llbra\alpha,\beta\rrket
    = -(-1)^{(a-1)(b-1)}\llbra\beta,\alpha\rrket
    = -(-1)^{(a-1)(b-1)}\S^\beta\alpha.
\end{equation*}
We refer the reader specifically to Ref.~\onlinecite{Huguet:2024mza} in which, among other things, these aspects of the Schouten--Nijenhuis bracket brought to differential forms are studied.

Let $\phi$ be a scalar field, key formulas for the present article are
\begin{align}
    &*^{-1}\L_{\sharp d\phi}* = \S^{d\phi} + (\square_g\phi),  \label{eq:*L*}\\
    &\S^{d\phi} = (\L_{\sharp d\phi}\tilde{g})^{ab}j_a i_b + \L_{\sharp d\phi},
    \label{eq:Sdphi0}
\end{align}
using Eqs.~(9) and (19) of Ref.~\onlinecite{Huguet:2024mza}.
For $\tilde{g}=\tilde{\eta}$ one has the simplification
\begin{equation*}
    (\L_{\sharp d\phi}\tilde{\eta})^{\alpha\beta} = -2(\dr^\alpha\dr^\beta\phi),
\end{equation*}
this is an instance of Lemma~1.60 of Ref.~\onlinecite{Besse:1987pua}, for which, in general, we have
\begin{equation}
    \L_{\sharp_gd\phi}\tilde{g}
    = -2\nabla^g d\phi. \label{eq:Lie=Hessien}
\end{equation}
That is to say that Eq.~\eqref{eq:Sdphi0} can be rewritten as
\begin{equation}
    \label{eq:Sdphi}
    \S^{d\phi} = (\nabla^g d\phi)^{ab}j_a i_b + \L_{\sharp d\phi}
    ={}^{\sss D}\nabla^g d\phi+ \L_{\sharp d\phi},
\end{equation}
where ${}^{\sss D}\nabla^g d\phi$ is the unique derivation on $\Omega(M)$ of order 0, vanishing on scalar fields and coinciding with the contraction with $\nabla^g d\phi$ when applied on a 1-form.

\subsection{Curvature and Kulkarni--Nomizu product}

We recall\cite{Fecko,jost2008riemannian} that,
for $\nabla^g$ the Levi-Civita connection, the curvature operator is defined as
\begin{equation*}
    R(u,v)w
    = (\nabla_u^g\nabla_v^g - \nabla_v^g\nabla_u^g + \nabla^g_{[u,v]})w
    = [(\nabla^g)^2_{u,v} - (\nabla^g)^2_{v,u}]w,
\end{equation*}
for $u$, $v$, and $w$ vectors, and $(\nabla^g)^2$ the second covariant derivative.
Out of the curvature operator we single out the Riemann tensor and its various traces,
\begin{align*}
    &\Riemann(x,w,u,v) = g(x,R(u,v)w)
    = g(e_c,R(e_a, e_b)e_d) x^c w^d u^a v^b
    = R_{cd ab} x^c w^d u^a v^b,\\
    &\Ricci(u,v) 
    = \tilde{g}^{cd} g(u,R(v,e_c)e_d)
    = \tilde{g}^{cd} R_{a cb d} u^a v^b
    = (C_{2,4}\Riemann)(u,v)
    = (C_{1,3}\Riemann)(u,v),\\
    &\R = \tilde{g}^{ab}\Ricci(e_a, e_b) = C_{1,2}\Ricci,
\end{align*}
using the symmetry of the Riemann tensor to shuffle the indices.
Here $C_{i,j}$ is the contraction between the $i$th and the $j$th entries.

\begin{remark}
    Note that the curvature operator, \emph{as an operator on $\Omega(M)$}, can be expressed in terms of the Riemann tensor  using 
    \begin{equation}
        \label{Eq: Riemann}
        R(u,v)=\Riemann(e_a,e_b,u,v)j^ai^b.
    \end{equation}
    Indeed, any derivation $T$ of order 0 on $\Omega(M)$ vanishing on scalar fields fulfills the relation $T=\langle Te^a,e_b\rangle j^bi_a$. In fact, since $\langle Te^a,e_b\rangle j^bi_a$ is also a derivation of order 0 vanishing on scalar fields, it suffices to verify this relation for the one-forms, which is straightforward.
    Then, applying this relation to $R(u,v)$, taking into account that $R(u,v)$ commutes with contractions, gives the result~\eqref{Eq: Riemann}.
\end{remark}

The formula relating the Laplace--de Rham and Laplace--Beltrami operators is the Weitzenböck formula.\cite{jost2008riemannian}
It can be written as
\begin{equation*}
    \square\alpha=\Delta\alpha+j^ai^bR(e_a,e_b)\alpha,
\end{equation*}
see the appendix of Ref.~\onlinecite{Huguet:2022rxi}.
Now, together with  Eq.~\eqref{Eq: Riemann}, it can be recast in a form useful to our purpose as
\begin{equation}
    \square\alpha=\Delta\alpha+\Riemann_{abcd}j^ai^bj^ci^d\alpha.
    \label{eq: Weitzenbock}
\end{equation}

The Kulkarni--Nomizu product,\cite{Besse:1987pua} denoted $\owedge$, is a product that from two symmetric type (0,2) tensors builds out a tensor of type (0,4) fulfilling the algebraic properties of the Riemann tensor.
For $h$ and $k$ two symmetric tensors and $x$, $w$, $u$, and $v$ vectors it is given by
\begin{equation}\label{eq:Def_KN}
    (h\owedge k)(x,w,u,v)
    = h(x,u)k(w,v)
    + h(w,v)k(x,u)
    - h(w,u)k(x,v)
    - h(x,v)k(w,u).
\end{equation}
One directly has that $h\owedge k = k\owedge h$ and that
\begin{equation*}
    (h\owedge h)(x,w,u,v)
    = 2[h(x,u)h(w,v)- h(w,u)h(x,v)].
\end{equation*}
Furthermore, for $g$ the metric on $M$ of dimension $n$, one has
\begin{align}
    &C_{1,3}(g\owedge k) = (n-2)k + \Tr(k) g,\label{eq:C_13}\\
    &C_{1,3}(g\owedge g) = 2(n-1)g. \label{eq:C_13g}
\end{align}

In the case of AdSM spaces, which are maximally symmetric and of constant curvature, one has
\begin{equation*}
    \Riemann_\kappa = -\frac{\kappa}{2} g\owedge g,\qquad
    \Ricci_\kappa = -\kappa(n-1) g,\qquad
    \R_\kappa = -n(n-1)\kappa,
\end{equation*}
with $\kappa$ being either $H^2$, 0 or $-H^2$, corresponding to de Sitter, Minkowski, or anti--de Sitter spaces.

\section{The rationale behind the embedding formulas of FLRW spaces (Prop.~\ref{prop:SimpleEmb})}
\label{app:Additional_Embedding}

We seek embedding formulas of FLRW spaces as simply as possible.

\subsection{Embeddings via Minkowski space as base space}

Since FLRW spaces are conformally flat there exist coordinate transformations such that
\begin{equation}\label{eq:EtaRW-Mink}
    \eta_{\sss{FLRW}} = \Omega^2(\xi)\eta_0(\xi),
\end{equation}
with $\eta_0$ the metric over Minkowski space, $\xi^\mu$ Cartesian coordinates, and $\Omega$ a function depending upon the scale factor $a$ and $\xi^\mu$.
With Minkowski space as base space one has that 
\begin{equation*}
    \left\{
    \begin{aligned}
        &y^\mu=\Omega(\xi)\xi^\mu,\\
        &y^n=\tfrac{1}{2}\Omega(\xi)(1 + \xi_\mu \xi^\mu),\\
        &y^{n+1}=\tfrac{1}{2}\Omega(\xi)(1 - \xi_\mu \xi^\mu),
    \end{aligned}
    \right.
\end{equation*}
is an embedding in $\setR^{n+2}$ of FLRW spaces.
The difficulty here resides in finding a transformation bringing the FLRW metric [Eq.~\eqref{eq:EtaRW}] in the form of Eq.~\eqref{eq:EtaRW-Mink} and in particular to express the conformal factor $\Omega$ in terms of Cartesian coordinates.

The case $k=0$ is simple.
Indeed, since in the usual Cartesian coordinates one has
\begin{equation}\label{eq:Mink_k=0}
    \eta_0 = d\xi^0\otimes d\xi^0 - g_{\mathbb{E}^{n-1}},
\end{equation}
with $g_{\mathbb{E}^{n-1}}$ the metric on the flat Euclidean space of dimension $n-1$,
we recognize that $\eta_0$ is a $k=0$ FLRW metric, thus
\begin{equation}
    \label{eq:Ak0}
    \Omega(\xi) = a(\xi^0).
\end{equation}
That is: $\Omega$ depends only on one variable $\xi^0$ through the scale factor $a$.
With this simple conformal factor $\Omega$ and the corresponding scaling in $\setR^{n+2}$ we obtain the embedding given in Eq.~\eqref{eq:SimpleEmbk=0} and the $k=
0$ case of Eq.~\eqref{eq:SimpleEmb} by setting $t=\xi^0$ and $\chi = \sqrt{\sum_i(\xi^i)^2}$.

For $k=\pm 1$ the conformal factor $\Omega$ is much more involved, we reproduce the result\cite{Iihoshi:2007uz,Ibison:2007dv}
\begin{equation}\label{eq:Akpm1}
    \Omega(\xi) =  \frac{1}{\sqrt{(1-\frac{1}{4}k\xi_\mu\xi^\mu)^2 + k(\xi^0)^2}}
        \,
        a\left(\frac{1}{\sqrt{k}}\tan^{-1}\left(
            \frac{\sqrt{k}\xi^0}{1-\frac{1}{4}k\xi_\mu\xi^\mu}
        \right)\right),
\end{equation}
depending on whether $k$ is positive or negative, one uses either $\tan^{-1}$ or $\tanh^{-1}$, and as $k$ approaches zero Eq.~\eqref{eq:Ak0} is retrieved.
In the case $k=-1$ of the FLRW metric, there is a specific additional ``exponential transformation'' where the conformal factor can be expressed in the form
\begin{equation}
    \label{eq:A_Exponential}
    \Omega(\xi) = \frac{1}{4\sqrt{\xi_\mu\xi^\mu}}\, a\left(
        \tfrac{1}{2}\ln(\xi_\mu\xi^\mu)\right).
    \qquad (k=-1)
\end{equation}
For detailed information on the transformations used to derive Eqs.~\eqref{eq:Akpm1} and \eqref{eq:A_Exponential} refer to the works in Refs.~\onlinecite{Iihoshi:2007uz} and \onlinecite{Ibison:2007dv}.
The aspect of the conformal factors in Eqs.~\eqref{eq:Akpm1} and \eqref{eq:A_Exponential} make them unfit for our pursuit of simple formulas, while the conformal factor in Eq.~\eqref{eq:Ak0} is exactly of the kind we are looking for.

\subsection{de Sitter space as an all-purpose base space}
\label{subsec:Improved_dS}

The property that made simple the embedding of FLRW spaces of type $k=0$ [Eqs.~\eqref{eq:SimpleEmb} and \eqref{eq:SimpleEmbk=0}] via Minkowski space is the fact that in Cartesian coordinates $\eta_0$ is \emph{also} a FLRW metric of type $k=0$.
Specifically, a successful approach to reach simple embeddings is to take a base space whose metric has been turned into an FLRW metric of the desired $k$ type.
Then, there necessarily exists a conformal factor $\Omega$ that relates the two metrics, which depends solely on the conformal time $t$ [as shown in Eq.~\eqref{eq:Ak0} for $k=0$].
De Sitter space possesses the unique feature\cite{Torrence_Couch_1986} that, using specific coordinate systems, its metric can assume each of the three FLRW forms.
Hence, with this property, dS space provides a better suited base space for the embedding of FLRW spaces in $\setR^{n+2}$.

\begin{figure}
    \centering
    \includegraphics{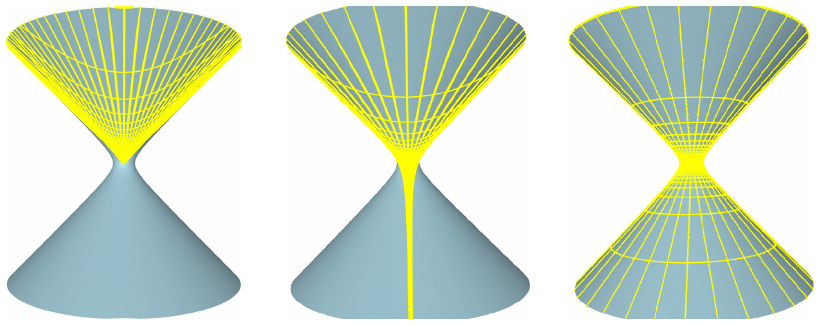}
    \caption{Coordinates $(t,\chi)$ on the de Sitter spacetime given in Eqs.~\eqref{eq:dS_k=-1}--\eqref{eq:dS_k=1} drawn on the embedded hyperboloid in $\setR^{n+1}\subset\setR^{n+2}$ (obtained by suppressing the $y^{n+1}$ coordinate and reducing the $y^i$s coordinates to a single coordinate).
    We sequentially represent the cases $k=-1$, $k=0$ and $k=+1$.}
    \label{fig:Coord_Sur_dS}
\end{figure}

First, we use various known coordinates systems\cite{Birrell:1982ix,Eriksen:1995ws,Spradlin:2001pw,Moschella:2006pkh} realizing the FLRW form of the dS metric
\begin{align}
    k=-1\ :&\ \left\{
        \begin{aligned}
            &y^0 = H^{-1}\csch(t)\cosh(\chi), \\
            &y^i = H^{-1}\csch(t)\sinh(\chi)\omega^i,\\
            &y^n = H^{-1}\coth(t),\\
            &y^{n+1} = H^{-1},
        \end{aligned}\right.
    && \eta_{\sss{H^2}} = H^{-2}\csch^2(t)\{dt \otimes dt - g_{\mathbb{H}^{n-1}}\},
        \label{eq:dS_k=-1}\\
    k=0\ :&\ \left\{
        \begin{aligned}
        &y^0 = \tfrac{1}{2}H^{-1}[-t + t^{-1} + t^{-1}\chi^2],\\
        &y^i = H^{-1}t^{-1}\chi\omega^i,\\
        &y^n = \tfrac{1}{2}H^{-1}[t + t^{-1} - t^{-1}\chi^2],\\
        &y^{n+1} = H^{-1},
        \end{aligned}
        \right.
    && \eta_{\sss{H^2}} = 
        (H t)^{-2}\{dt\otimes dt - g_{\mathbb{E}^{n-1}}\},
        \label{eq:dS_k=0}\\
    k=+1\ :&\ \left\{
        \begin{aligned}
        &y^0=H^{-1}\cot(t),\\
        &y^i=H^{-1}\csc(t)
            \sin(\chi)\omega^i,\\
        &y^n=H^{-1}\csc(t)\cos(\chi),\\
        &y^{n+1}=H^{-1},
        \end{aligned}
    \right.
    && \eta_{\sss{H^2}} =
            H^{-2}\csc^2(t)\left\{
            dt\otimes dt - g_{\mathbb{S}^{n-1}}\right\},
        \label{eq:dS_k=1}
\end{align}
where, in each case, $\omega^i$ are ambient coordinates on $\mathbb{S}^{n-2}$ [i.e., $\sum_i (\omega^i)^2 = 1$], the variable $\chi$ is related to the radial coordinate $r$ of the isotropic form of the spatial metrics by $r = \sin(\sqrt{k}\chi)/\sqrt{k}$, and the different spatial metrics $g_{\mathbb{H}^{n-1}}$, $g_{\mathbb{E}^{n-1}}$, and $g_{\mathbb{S}^{n-1}}$ are the metrics on the $n-1$ dimensional hyperboloid, Euclidean space, and sphere, respectively.
See Fig.~\ref{fig:Coord_Sur_dS} for a depiction of each coordinate system.

Then, the FLRW metric [Eq.~\eqref{eq:EtaRW}] can be made explicitly conformally related to the dS metric by writing
\begin{equation}
    \label{eq:Ak_dS}
    \eta_{\sss{FLRW}} = \Omega^2(t)\eta_{\sss{H^2}}\quad
    \text{with}\quad
    \Omega(t) = \begin{cases}
        a(t)H\sinh(t) &\text{for $k=-1$,}\\
        a(t)Ht &\text{for $k=0$,}\\
        a(t)H\sin(t) &\text{for $k=+1$}.
    \end{cases}
\end{equation}
Finally, through the scaling in $\setR^{n+2}$ by $\Omega(t)$ of Eqs.~\eqref{eq:dS_k=-1}--\eqref{eq:dS_k=1}, we obtain the embedding formulas of Eq.~\eqref{eq:SimpleEmb} for $k=\pm 1$ and the $k=0$ formula of Eq.~\eqref{eq:SimpleEmb} with $y^0$ and $y^{n+1}$ interchanged.
Additionally, note that while the systems presented in Eqs.~\eqref{eq:dS_k=-1}--\eqref{eq:dS_k=1} are singular for $t=0$ this singularity is absent in Eq.~\eqref{eq:SimpleEmb} when embedding the FLRW spaces into $\setR^{n+2}$, provided $a(t)$ does not add singularities.

\subsection{Minkowski and AdS spaces as alternative base spaces for $k=-1$}

In its Cartesian coordinates $\xi$ the metric $\eta_0$ over Minkowski space is naturally in the $k=0$ form of the FLRW metric.
Additionally, it can be turned into a $k=-1$ form of the FLRW metric\cite{Torrence_Couch_1986} as
\begin{equation}
    \label{eq:Mink_k=-1}
    \eta_0 = e^{-2t}\{dt\otimes dt - g_{\mathbb{H}^{n-1}}\}
\end{equation}
by considering the change to the variables 
\begin{equation*}
    t = -\tfrac{1}{2}\ln\left(\xi_\mu\xi^\mu\right),\quad
    \chi = \tfrac{1}{2}\ln\left(\frac
        {\xi^0 - \sqrt{\sum_i (\xi^i)^2}}{\xi^0 + \sqrt{\sum_i (\xi^i)^2}}\right),
\end{equation*}
which can be found to be related to the ``exponential transformation'' valid for $k=-1$, see Eq.~\eqref{eq:A_Exponential}.
Now, for $k=-1$, one has
\begin{equation*}
    \eta_{\sss{FLRW}} = \Omega^2(t)\eta_0\quad
    \text{with}\quad
    \Omega(t) = a(t) e^{t}.
\end{equation*}
Then, by scaling coordinates on Minkowski space in $\setR^{n+2}$ [Eq.~\eqref{eq:Coord_Mink}] one gets the embedding formula for the $k=-1$ case of the FLRW metric
in Eq.~\eqref{eq:SimpleEmb}, but obtained via Minkowski space.

Finally, and for completeness, the metric on the AdS space can (only) take the $k=-1$ FLRW form using the following coordinates
\begin{align}
    k=-1\ :&\ \left\{
        \begin{aligned}
        &y^0=H^{-1}\sech(t)\cosh(\chi),\\
        &y^i=H^{-1}\sech(t)\sinh(\chi)\omega^i,\\
        &y^n=H^{-1},\\
        &y^{n+1}=H^{-1}\tanh(t),
        \end{aligned}
    \right.
    && \eta_{\sss{-H^2}} =
            H^{-2}\sech^2(t)\left\{
            dt\otimes dt - g_{\mathbb{H}^{n-1}}\right\}.
        \label{eq:AdS_k=-1}
\end{align}
Now, one can relate the $k=-1$ FLRW metric to the AdS metric in its $k=-1$ FLRW form with
\begin{equation*}
    \eta_{\sss{FLRW}} = \Omega^2(t)\eta_{\sss{-H^2}}\quad
    \text{with}\quad
    \Omega(t) = a(t)H\cosh(t).
\end{equation*}
By scaling the coordinates described in Eq.~\eqref{eq:AdS_k=-1} with the factor $\Omega(t)$, we retrieve the FLRW embedding of the $k=-1$ case.
This serves as a third method, via AdS space, in obtaining the $k=-1$ embedding given in Eq.~\eqref{eq:SimpleEmb}.

\begin{remark}
    Note that the forms given in Eqs.~\eqref{eq:Mink_k=0}, \eqref{eq:dS_k=-1}, \eqref{eq:dS_k=0}, \eqref{eq:dS_k=1}, \eqref{eq:Mink_k=-1}, and \eqref{eq:AdS_k=-1} are precisely the only six FLRW metrics that can be turned into the static form.
    See Ref.~\onlinecite{Florides_1980} for further details and explicit change of coordinates to the static case.
\end{remark}

\section{Isometries of FLRW spaces from $\setR^{n+2}$}
\label{sec:IsoFLRW}

In this section, we derive all isometries of the FLRW spaces utilizing our ambient space approach within $\setR^{n+2}$.
Our method not only simplifies the process but also allows for a systematic recovery of these isometries.
Let us set $J = J^{\alpha\beta}y_\alpha\dr_\beta$, with $J^{\alpha\beta} = -J^{\alpha\beta}$ constants, spanning the $o(2,n)$ algebra of the conformal group.
Seeking isometries of $X_f$ amounts to find all the (independent) $J$ such that $J(f) = 0$.

\begin{remark}
    In the following developments the scale factor $a$ is always determined up to an overall multiplicative numerical real constant which, for pure convenience, we set to $1$.
\end{remark}

\subsection{FLRW spaces with $k=-1$}

For the $k=-1$ case, whose defining function is given in Eq.~\eqref{eq:f_FLRW}, one has
\begin{equation*}
    J(f_{k=-1}) = \frac{1}{a^2\sqrt{(y^n)^2 -(y^{n+1})^2}}[
        (a y^n + \dot{a} y^{n+1})y_\alpha J^{\alpha n}
        -(a y^{n+1} + \dot{a} y^n) y_\alpha J^{\alpha n+1}
    ],
\end{equation*}
with $\dot a$ the derivative of the scale factor $a$.
Obviously, the $\tfrac{1}{2}n(n-1)$ generators $y_\mu\dr_\nu - y_\nu\dr_\mu$ of $SO(1,n-1)$ fulfill the equation $J(f)=0$, as it was already remarked upon from the embedding formulas Eq.~\eqref{eq:SimpleEmb}.
We are interested in the additional isometries.
First, we note that $y_n\dr_{n+1} - y_{n+1}\dr_n$ is a solution to $J(f) = 0$ if $\dot{a} = 0$, that is $a=1$ and the space is the $k=-1$ Einstein space.
Regarding the remaining possible isometries it suffices to inspect the case with the index $\alpha = 0$, which reads as
\begin{equation*}
    J(f_{k=-1}) = \frac{y_0y^n}{a^2(t)\sqrt{(y^n)^2 -(y^{n+1})^2}}\{
        [a(t) + \dot{a}(t) \tanh(t)] J^{0 n}
        -[a(t) \tanh(t) + \dot{a}(t)] J^{0 n+1}
    \},
\end{equation*}
with $y^{n+1}/y^n = \tanh(t)$.
Then, $J(f_{k=-1}) = 0$ splits into three cases
    (i) $J^{0n} = 0$ and $J^{0n+1}\neq 0$,
    (ii) $J^{0n} \neq 0$ and $J^{0n+1} = 0$,
    (iii) $J^{0n}\neq 0$ and $J^{0n+1} \neq0$.
In the first case, one reaches an isometry if $\dot{a} + \tanh(t) a = 0$ that is for $a(t) = 1/\cosh(t) = \sech(t)$ thus recovering an AdS space [Eq.~\eqref{eq:AdS_k=-1}] where the additional isometries are $y_\mu\dr_{n+1} - y_{n+1}\dr_\mu$ completing $SO(1,n-1)$ to $SO(2,n-1)$.
The second case is very similar to the first case, one obtains additional isometries if $\dot{a} + a/\tanh(t) =0$ that is for $a(t) = 1/\sinh(t) = \csch(t)$ thus recovering a dS space [Eq.~\eqref{eq:dS_k=-1}] where the additional isometries are $y_\mu\dr_n - y_n\dr_\mu$ completing $SO(1,n-1)$ to $SO(1,n)$.
For the final case, let us set $q = J^{0n+1}/J^{0n}$ a nonzero real constant, then reaching an additional isometry amounts to solve
\begin{equation}
    \label{eq:dotA_k=-1}
    \dot{a}(t) + \left(\frac{1 - q\tanh(t)}{\tanh(t) - c}\right)a(t) = 0.
\end{equation}
Depending on $q$, Eq.~\eqref{eq:dotA_k=-1} decomposes into three sub-cases.
For $|q| < 1$, setting $q = \tanh(t_0)$, Eq.~\eqref{eq:dotA_k=-1} can be rewritten as $\dot a + a/\tanh(t-t_0) = 0$, whose solution is another dS space whose conformal time origin has been offset.
Similarly, for $|q| > 1$, setting $1/q = \tanh(t_0)$ the equation can be rewritten as $\dot a + \tanh(t-t_0) a = 0$, whose solution is another AdS space whose conformal time origin has been offset.
So, these two cases are nothing but case (i) and (ii) in a different guise.
The only original case happens for $q=\pm 1$ in which $\dot{a} = \pm a$, whose solution is $a(t) = e^{\pm t}$ corresponding to a Minkowski space [Eq.~\eqref{eq:Mink_k=-1}].
Thus, choosing the negative value to exactly match Eq.~\eqref{eq:Mink_k=-1}, one can add to the Lorentz subgroup $SO(1,n-1)$ the additional (ambient) generators $P_\mu = \tfrac{1}{2}[y_\mu(\dr_n -\dr_{n+1}) - (y_n - y_{n+1})\dr_\mu]$ of translations.
This analysis exhausts all the additional isometries that can be reached in the $k=-1$ case of FLRW spaces.

\subsection{FLRW spaces with $k=0$}

For the $k=0$ case, whose defining function is given in Eq.~\eqref{eq:f_FLRW}, one has
\begin{equation*}
    J(f_{k=0}) = \frac{1}{a^2}\{
        [a(t) + t\dot{a}(t)] y_\alpha( J^{\alpha n} + J^{\alpha n+1})
        + \dot{a}(t) y_\alpha J^{0\alpha}
    \},
\end{equation*}
with $t=y^0/(y^n+y^{n+1})$.
The equation $J(f) = 0$ is obviously fulfilled by the $\tfrac{1}{2}(n-1)(n-2)$ rotations generated by $y_i\dr_j - y_j\dr_i$ and by the $(n-1)$ translations $P_i = \tfrac{1}{2}[y_i(\dr_n -\dr_{n+1}) - (y_n - y_{n+1})\dr_i]$, together they generate the (minimal) group of isometries $E(n-1)$ of the $k=0$ FLRW spaces.
Finding all of the solutions to the above equation is again rather straightforward.
First by considering the case of $\alpha = i$ one has three cases to consider:
    (i) $J^{0i} = 0$ and $(J^{in} + J^{in+1})\neq 0$,
    (ii) $J^{0i} \neq 0$ and $(J^{in} + J^{in+1}) = 0$,
    (iii) $J^{0i} \neq 0$ and $(J^{in} + J^{in+1}) \neq 0$.
In the first case, one has additional isometries if and only if $a$ is a solution to $t\dot{a} + a = 0$, that is for $a(t) = t^{-1}$, which corresponds to a dS space [Eq.~\eqref{eq:dS_k=0}].
To the generators of $E(n-1)$ one can then add the $(n-1)$ extra generators $T_i = \tfrac{1}{2}[y_i(\dr_n + \dr_{n+1}) - (y_n+y_{n+1})\dr_i]$ and the hyperbolic rotation $y_n\dr_{n+1} - y_{n+1}\dr_n$.
Thus, considering then $P_i\pm T_i$, we recover the full de Sitter group $SO(1,n)$ with the generators $y_i\dr_j-y_j\dr_i$, $y_n\dr_i-y_i\dr_n$, $y_{n+1}\dr_i - y_i\dr_{n+1}$ and $y_n\dr_{n+1}-y_{n+1}\dr_n$, that is the usual ones in which the roles of $y^0$ and $y^{n+1}$ have been swapped.
In the second case, one has additional isometries if $\dot{a} = 0$, that is $a=1$ and the FLRW space simply is Minkowski space [Eq.~\eqref{eq:Mink_k=0}].
Then, to $E(n-1)$ one can add the $n-1$ hyperbolic rotations $y_0\dr_i - y_i\dr_0$ and the translation $P_0 = \frac{1}{2}[y_0(\dr_n - \dr_{n+1}) - (y_n -y_{n+1})\dr_0]$ thus obtaining the Poincaré group.
For the final case, we set $t_0 = -J^{0i}/(J^{in} + J^{in+1})$ and the equation to solve to obtain additional isometries is $(t-t_0)\dot{a}(t) + a(t) = 0$, whose solution is another dS space, which we studied in case (i).
Now, for $\alpha\neq i$, one has to find the most general solutions to the system
\begin{equation*}
    \left\{\begin{aligned}
        &[a(t) + t\dot{a}(t)](t (J^{0n} + J^{0n+1}) - J^{nn+1})
            -\dot{a}(t) J^{0n} = 0,\\
        &[a(t) + t\dot{a}(t)](t (J^{0n} + J^{0n+1}) - J^{nn+1})
            +\dot{a}(t) J^{0n+1} = 0,
    \end{aligned}\right.
\end{equation*}
using the fact that $y^0 = t (y^n + y^{n+1})$ and that the vanishing of $J(f)$ has to hold $\forall y^n, y^{n+1}$.
First we discard the solutions to $a(t) + t\dot{a}(t) = 0$ and $\dot{a}(t) = 0$ corresponding, respectively, to dS space and Minkowski space, which we solved already just above.
Under this assumption, by taking the difference of the two equations we have to have $(J^{0n}+J^{0n+1}) = 0$, using this in the sum of the two equations we get the differential equation $[a(t) + t \dot{a}(t)] J^{nn+1} + \dot{a}(t) (J^{0n+1} - J^{0n}) = 0$, which is nothing but the case (iii) solved previously, corresponding to a dS space.
That is, all the additional isometries to those of $E(n-1)$ were found in the cases (i) and (ii).

\subsection{FLRW spaces with $k=+1$}

For the $k=+1$ case, as its defining function resembles that of the $k=-1$ case, one expects to draw similar conclusions.
This is only partially the case.
The computations parallels completely those of the $k=-1$ case, which we skip here, focusing instead on the differences.
Very similarly to the $k=-1$ case we have one additional isometry $y_0\dr_{n+1} - y_{n+1}\dr_0$ if $\dot{a} = 0$, that is $a=1$ and the space is the $k=+1$ Einstein space.
Much like the $k=-1$ scenario, there exist three specific instances that result in the occurrence of $n$ extra isometries.
First, if $a(t)$ fulfills $\tan(t)\dot{a} + a = 0$, that is for $a(t) = 1/\sin(t) = \csc(t)$, corresponding to a dS space [Eq.~\eqref{eq:dS_k=1}].
Second, if $a(t)$ fulfills $\dot{a} - \tan(t) a = \dot{a} + a/\tan(t-\pi/2) = 0$, which is yet another instance of a de Sitter space.\footnote{
Otherwise a direct integration yields $a(t) = \cos(t)$ leading to the metric 
    $\eta_{\sss{H^2}}
    = \cos^2(t)[dt\otimes dt - g_{\mathbb{S}^{n-1}}]
    = d\tau\otimes d\tau - (1-\tau^2)g_{\mathbb{S}^{n-1}}$ over de Sitter space.}
Third, and finally, the equivalent to Eq.~\eqref{eq:dotA_k=-1} here reads as
\begin{equation*}
    \dot{a}(t) + \left(\frac{1+q\tan(t)}{\tan(t)-q}\right) a(t)
    = \dot{a}(t) + \frac{1}{\tan(t-t_0)}a(t)
    = 0,
\end{equation*}
with $q = \tan(t_0) = J^{in+1}/J^{i0}\neq 0$ and no sub-cases depending on $q$, whose solution is $a(t) = \csc(t-t_0)$ which, again, corresponds to a dS space.
That is, in the $k=+1$ case of the FLRW spaces, all the cases merge only into a dS space.

\subsection{Final summary}

In all cases, to the manifest (minimal) $\tfrac{1}{2}n(n-1)$ isometries of the FLRW spaces one has
\begin{itemize}
    \item one additional isometry if $X_{f_k}$ is an Einstein space ($k = \pm 1)$,
    \item $n$ additional isometries if $X_{f_k}$ is a de Sitter space ($k= \pm 1$, $k=0$),
        a Minkowski space ($k=-1$, $k=0$), or an anti--de Sitter space ($k=-1$).
\end{itemize}
These are the only eight cases with enhanced symmetry.
These are know facts\cite{Maartens_Maharaj_1986,Keane:1999kp} but set in the ambient space $\setR^{n+2}$ with a choice of basis in $o(2,n)$ reflecting our choice of $f$ leading to the space $X_f$.
This constitutes, also, a roundabout way to reach the fact that dS space admits the three forms of the FLRW metric, Minkowski space two forms, and AdS space only one.\cite{Torrence_Couch_1986}
The remaining elements of $o(2,n)$ that are not isometries then act as conformal killing vectors on $X_f$.\cite{Maartens_Maharaj_1986,Keane:1999kp}
A complete classification of conformally flat spaces by their minimal group of isometries is available in Ref.~\onlinecite{Backovsky_Niederle_1997}.
Minkowski, (anti--)de Sitter and FLRW\footnote{Einstein spaces are viewed as FLRW spaces in this classification.} spaces being the most symmetric among a list of 15 cases.

\section{Proofs of various properties}
\label{sec:Proofs}

In this appendix we gather various proofs whose length could impede the reading of the main body of the article.

\subsection{Proof of Prop.~\ref{prop:Restriction}}
\label{proof:Restriction}

In order to prove Prop.~\ref{prop:Restriction} it has proven useful, for $y\in X_f$, to use
\begin{align*}
    &T = \iF\iD j^{df}j^{dc},\\
    &T_c = *_\eta^{-1}T*_\eta =  j^{df}j^{dc}\iF\iD,
\end{align*}
which, for $y\in X_f$, fulfill
\begin{equation*}
    T^2 = T,\quad
    (T_c)^2 = T_c,\quad
    TT_c = T_cT = 0.
\end{equation*}
Still for $y\in X_f$ with the representation of the unity
\begin{align*}
    \Id &= T - T_c + j^ni_n + j^{n+1}i_{n+1}\\
    &= T_c - T + i_nj^n + i_{n+1}j^{n+1},
\end{align*}
introducing the identity  within $T$ and $T_c$ yields
\begin{align*}
    &T = -i_ni_{n+1}\Id j^n j^{n+1} = -i_ni_{n+1}T_c j^n j^{n+1} = \iF\iD T_c j^{df}j^{dc},\\
    &T_c = -j^nj^{n+1}\Id i_ni_{n+1} = -j^nj^{n+1}T i_ni_{n+1} = j^{df}j^{dc}T \iF\iD.
\end{align*}

\begin{proof} (of Prop.~\ref{prop:Restriction})
    Let us begin with Eq.~\eqref{eq:Restr_*}, which follows from a straightforward computation involving $T$ and $T_c$.
    Indeed,
    \begin{align*}
        m_f{}^*(*_\eta\alpha)
        &= m_f{}^*(T^2*_\eta\alpha)
        = m_f{}^*(T*_\eta T_c\alpha)
        = -m_f{}^*(T*_\eta j^nj^{n+1}T i_ni_{n+1}\alpha)\\
        &= -m_f{}^*(T*_\eta j^nj^{n+1}T) m_f{}^*(i_ni_{n+1}\alpha)
        = *_f  m_f{}^*(i_ni_{n+1}\alpha)
        = *_f m_f{}^*(\iF\iD\alpha),
    \end{align*}
    with then $*_f = -m_f{}^*(*_\eta j^nj^{n+1}) = m_f{}^*(i_{n+1}i_n*_\eta)$
    and in particular $*_f 1= m_f{}^*(i_{n+1}i_n *_\eta 1) = \omega_f$, as it should.

    Equation~\eqref{eq:Restr_d} is well known (see Eq.~6.2.12 of Ref.~\onlinecite{Fecko}).
    
    Regarding Eq.~\eqref{eq:Restr_delta} one has
    \begin{align}
        m_f{}^*(\delta\alpha)
        &= \sgn(\eta) (-1)^{(n+2)(a+1)+1}m_f{}^*(*_\eta d*_\eta\alpha) \nonumber\\
        &= \sgn(\eta) (-1)^{n(a+1) + 1}*_fm_f{}^*(\iF\iD d *_\eta\alpha) \nonumber\\
        &= \sgn(\eta) (-1)^{n(a+1) + 1}\{*_fd_fm_f{}^*(\iF\iD*_\eta\alpha)
        + *_fm_f{}^*[(\iF(\LD -1) - \iD\LF)*_\eta\alpha]\},       \label{eq:mfDeltaA_1}
    \end{align}
    where we used
    \begin{equation*}
        \iF\iD d
        = d\iF\iD + \iF\LD - \LF\iD
        = d\iF\iD + \iF(\LD -1) - \iD\LF,
    \end{equation*}
    with $[\LF,\iD] = \iF$.
    The first term of Eq.~\eqref{eq:mfDeltaA_1} simply is $\delta_f\alpha_f$ since $m_f{}^*(\iF\iD*_\eta\alpha) = -*_f\alpha_f$ and that $\sgn(\eta) = -\sgn(\eta_f)$, balancing the signs.
    Now, regarding the second term, one is led to consider
    \begin{align}
        *_fm_f{}^*&[(\iF(\LD -1) - \iD\LF)*_\eta\alpha]
        = -m_f{}^*[\iF\iD*_\eta  \{\iF(\LD - 1) -\iD\LF\}*_\eta\alpha]\nonumber\\
        &= -\sgn(\eta)(-1)^{n(a+1)+1}m_f{}^*[
            \iF\iD \{j^{df}(\S^{dc} + \square_\eta c - 1)
            - j^{dc}(\S^{df} + \square_\eta f)\}\alpha],
            \label{eq:Terme_II}
     \end{align}
    using Eq.~\eqref{eq:*L*} and the fact that, for $\beta\in\Omega^b(\setR^{n+2})$, $*_\eta *_\eta\beta = \sgn(\eta)(-1)^{(n+1)b}\beta$.
    Note that the sign in Eq.~\eqref{eq:Terme_II} cancels with the overall sign of the second term in Eq.~\eqref{eq:mfDeltaA_1}.
    This previous equation can be further simplified with
    \begin{align*}
        &m_f{}^*(\iF\iD j^{df}\beta)
        = m_f{}^*[(f\iF - \Fd\iD)\beta]
        = m_f{}^*[(\iF - \Fd\iD)\beta],\\
        &m_f{}^*(\iF\iD j^{dc}\beta)
        = m_f{}^*[(2c\iF - f\iD)\beta]
        = -m_f{}^*(\iD\beta),\\
        &\S^{dc}\beta 
        = [(\LD\tilde\eta)^{\alpha\beta}j_\alpha i_\beta + \LD]\beta 
        = (-2\tilde{\eta}^{\alpha\beta}j_\alpha i_\beta + \LD)\beta
        = (\LD - 2b)\beta,
    \end{align*}
    since $(\dr^\alpha\dr^\beta c) = \tilde\eta^{\alpha\beta}$ and $\square_\eta c = n+2$
    one gets that the additional term to $\delta_f\alpha_f$ is
    \begin{align*}
        -m_f{}^*&[\{(\iF - \Fd\iD)(\LD -2a + n +1)
            + \iD(\S^{df}+\square_\eta f)\}\alpha]\\
            &=-m_f{}^*[\{(\LD -2a +n+1)\iF 
                - \Fd(\LD -2a +n +1)\iD 
                + (\S^{df} + \square_\eta f)\iD\}\alpha],
    \end{align*}
    since $\iF\LD = (\LD+1)\iF$,
    $(\LF dc) = dF(c) = df$
    and using the identity
    \begin{equation*}
        \iD(\S^{df} + \square_\eta f)\alpha
        = \iD *_\eta^{-1}\LF*_\eta\alpha
        = [(\S^{df}+\square_\eta f)\iD - \iF]\alpha.
    \end{equation*}
    This establishes Eq.~\eqref{eq:Restr_delta}.

    Finally, regarding the pull-back of the Laplacian $\square_\eta = -(d\delta_\eta + \delta_\eta d)$, given in Eq.~\eqref{eq:Restr_Box}, according to Eqs.~\eqref{eq:Restr_d} and \eqref{eq:Restr_delta} its leading term is $\square_f\alpha_f$ with a first additional term
    \begin{equation*}
        +m_f{}^*[(\LD -2a +n+1)\LF\alpha 
        + (\S^{df} + \square_\eta f - \Fd(\LD -2a +n +1))\LD\alpha],
    \end{equation*}
    and a second set of additional terms coming from $(d\Fd)\wedge\beta = j^{d\Fd}\beta$ and from the commutator
    $[d,(\S^{df} + \square_\eta f)]\iD\alpha$.
\end{proof}
\begin{remark}
    Notice that this peculiar term can be rewritten using
    \begin{align*}
        [d,(\S^{df} + \square_\eta f)]\beta
        &= [d,*_\eta^{-1}\LF*_\eta]\beta \\
        &= ((-1)^{n+1}-1)*_\eta^{-1}\LF*_\eta d\beta
            +(-1)^{n-b}*_\eta^{-1}[\delta_\eta,\LF]*_\eta\beta   \\
        &= ((-1)^{n+1}-1)*_\eta^{-1}\LF*_\eta d\beta
            +(-1)^{n-b}*_\eta^{-1}[\square_\eta,\iF]*_\eta\beta  \\
        &= ((-1)^{n+1}-1)(\S^{df} + \square_\eta f) d\beta
            +(-1)^{n}[\square_\eta,j^{df}]\beta,
    \end{align*}
    using the commutation relation between the codifferential and the Lie derivative [see Eq.~(80) of Ref.~\onlinecite{Huguet:2024mza}], which simplifies here since $d\flat F= ddf= 0$
    and with the nice fact that $*^{-1}\square *\beta = \square\beta$.
    Then, if need be, the corresponding additional term in Eq.~\eqref{eq:Restr_Box} can be simplified to
    \begin{equation*}
        +m_f{}^*[((-1)^{n+1}-1)(\S^{df} + \square_\eta f) d\iD\alpha +
            (-1)^n\square_\eta j^{df}\iD\alpha].
            \qedhere
    \end{equation*}
\end{remark}

\subsection{Proof of Prop.~\ref{prop:Hessien}}\label{proof:Hessien}

\begin{proof}
   Let $u,v\in T(X_f)$ and $\hat{u}$ and $\hat{v}$ be extensions to the ambient space.
   The very definition of $m_f{}^*$ implies that 
   $m_f{}^*(\nabla^\eta d\phi)(u,v)=(\nabla^\eta d\phi(\hat u,\hat v))_{|X_f}$.
    Then
    \begin{equation*}
        (\nabla^\eta d\phi)(\hat{u},\hat{v})
        = (\nabla_{\hat{v}}^\eta d\phi)(\hat{u})
        = \langle\nabla_{\hat{v}}^\eta d\phi, \hat{u}\rangle
        = \eta(\sharp_\eta \nabla_{\hat{v}}^\eta d\phi, \hat{u})
        = \eta(\nabla_{\hat{v}}^\eta \sharp_\eta d\phi, \hat{u}).
    \end{equation*}
    At this point it is important to split $\sharp_\eta d\phi$ as
    $\sharp_\eta d\phi = (\sharp_\eta d\phi)_\parallel + (\sharp_\eta d\phi)_\perp$,
    with $(\sharp_\eta d\phi)_\perp\in\mathrm{span}(D,F)$ and $(\sharp_\eta d\phi)_\parallel\in\mathrm{span}(D,F)^\perp$.
    
    Regarding the parallel part note that 
    \begin{align*}
        \eta(\nabla_{\hat{v}}^\eta (\sharp_\eta d\phi)_\parallel, \hat{u})\rvert_{X_f}
        &= [\hat{v}(\eta( (\sharp_\eta d\phi)_\parallel, \hat{u}))
            -\eta((\sharp_\eta d\phi)_\parallel, \nabla_{\hat{v}}^\eta\hat{u})
            ]\rvert_{X_f}\\
        &= [v(\eta_f(\sharp_f d_f\phi, u))
            -\eta_f(\sharp_f d_f\phi, \nabla_v^f u)]\rvert_{X_f}\\
        &= (\nabla_f d_f\phi_f)(u,v),
    \end{align*}
    using the fact that $\nabla^\eta$ is metric and Eq.~(4.7.1) of Ref.~\onlinecite{jost2008riemannian} stating, in our notations, that
    $\nabla^f_v u = (\nabla^\eta_v u)^{\mathsf{T}}$ with $\mathsf{T}$ the orthogonal projection $T_y(\setR^{n+2})\to T_y(X_f)$.
    Thus the parallel part of $\sharp_\eta d\phi$ yields the expected term: the Hessian of $\phi_f$.

    Regarding the normal part it is key to remark that around $X_f$ that $(\sharp_\eta d\phi)_\perp$ can be rewritten as
    \begin{align*}
        (\sharp_\eta d\phi)_\perp
        &= \sharp_\eta[e_n(\phi) e^n + e_{n+1}(\phi) e^{n+1}]\\
        &= D(\phi) F + (F(\phi) -\Fd D(\phi))D
        = D(\phi) F + E_f(\phi) D.
    \end{align*}
    Therefore, restricted to $X_f$ it leads to
    \begin{align*}
        \eta(\nabla_{\hat{v}}^\eta (\sharp_\eta d\phi)_\perp, \hat{u})\rvert_{X_f}
        &= [D(\phi) \eta(\nabla_{\hat{v}}^\eta F, \hat{u})
        + E_f(\phi) \eta(\nabla_{\hat{v}}^\eta D, \hat{u})]\rvert_{X_f}\\
        &= [\tfrac{1}{2} D(\phi) (\LF\eta)(\hat{u},\hat{v})
        + \tfrac{1}{2} E_f(\phi) (\LD\eta)(\hat{u},\hat{v})]\rvert_{X_f}\\
        &= D(\phi) (\nabla^\eta df)(\hat{u},\hat{v})
        + E_f(\phi) \eta(\hat{u},\hat{v}),
    \end{align*}
    using in the first line the fact that restricted to $X_f$ one has $\eta(F,\hat{u})\rvert_{X_f} = \eta(D,\hat{u})\rvert_{X_f} = 0$,
    with the fact that the Hessian is symmetric in $\hat{u}$, $\hat{v}$ and
    \begin{equation*}
        \frac{1}{2}[g(\nabla_u w, v) + g(\nabla_v w, u)]
        = \frac{1}{2}(\Lie_w g)(u,v),
    \end{equation*}
    which is a consequence of the identity $\L_w(g(u,v)) = \nabla_w(g(u,v))$.
    Then, the expression is simplified using the fact that $\LD\eta = 2\eta$ and that $\LF\eta = 2 (\nabla^\eta d f)$, 
    using Eq.~\eqref{eq:Lie=Hessien} on $g$ instead of $\tilde{g}$.
\end{proof}

\subsection{Proof of Prop.~\ref{prop:Extension}}
\label{sec:Extension}

\begin{proof}
First, from the fact that $\hat\beta$ is strongly transverse of degree zero Eqs.~\eqref{eq:Restr_delta} and \eqref{eq:Restr_Box} become
\begin{align*}
    &\delta_f\beta = m_f{}^*\delta_\eta\hat\beta, \\
    &\square_f\beta = m_f{}^*\square_\eta\hat\beta,
\end{align*}
which selects the transverse part of $\delta_\eta\hat\beta$ and of $\square_\eta\hat\beta$.
There remains to compute their longitudinal part.
In order to do so, we recall that, on $X_f$, the longitudinal part of a differential form $\alpha$ is picked out by 
\begin{align*}
    &m_f{}^l\alpha %
        =\bigl[(j^{df}\iD + j^{dc}\iF - \Fd j^{dc}\iD + j^{dc} j^{df}\iF\iD)\alpha\bigr]_{|X_f}.
\end{align*}

Note that $\iD\delta_\eta\hat\beta = - \delta_\eta\iD\hat\beta = 0$, by using Eq.~\eqref{eq:springboard1} to commute $\iD$ and $\delta_\eta$, with $d\flat D = d(dc) = 0$ [see Eq.~(62) of Ref.~\onlinecite{Huguet:2024mza} for the general formula], and then that $\hat\beta$ is strongly transverse.
Similarly we have that $\iF\delta_\eta\hat\beta = -\delta_\eta\iF\hat\beta = 0$.
Taking into account these anti-commutations we immediately obtain that $m_f{}^l(\delta_\eta\hat\beta) = 0$ and then that, at any point of $X_f$, we have
$\delta_\eta\hat\beta 
    = m_f{}^*\delta_\eta\hat\beta
    + m_f{}^l\delta_\eta\hat\beta
    = m_f{}^*\delta_\eta\hat\beta$, that is
\begin{equation*}
    \delta_f\beta = \delta_\eta\hat\beta,
\end{equation*}
which proves the first formula [Eq.~\eqref{eq:delta_fbeta}].

To compute the longitudinal part of $\square_\eta\hat\beta$ we first note that
\begin{align*} 
    \iF\square_\eta\hat\beta
    &=- \iF (d\delta_\eta + \delta_\eta d)\hat\beta
    =(d \iF -\LF)\delta_\eta\hat\beta + \delta_\eta \iF d\hat\beta\\
    &=-\LF\delta_\eta\hat\beta - d\delta_\eta\iF\hat\beta
    +\delta_\eta (-d\iF+\LF)\hat\beta
    =-\LF\delta_\eta\hat\beta,
\end{align*}
using $\iF d = \LF - d\iF$, the previous anti-commutation rule $\iF\delta_\eta = - \delta_\eta \iF$, and the fact that $\hat\beta$ is strongly transverse [see Eq.~(80) of Ref.~\onlinecite{Huguet:2024mza} for a generic formula regarding the commutation of the Lie derivative and $\delta$].
Similarly, we have
\begin{equation*}
    \iD\square_\eta\hat\beta
    = - \LD\delta_\eta\hat\beta %
    = 2 \delta_\eta\hat\beta,
\end{equation*}
where we used additionally the fact that %
$\delta_\eta$ is homogeneous of degree $-2$ (see Eq.~5.8.3 of Ref.~\onlinecite{Fecko} regarding the rescaling of the Hodge operator, or use the explicit expression of $\delta$ in Euclidean coordinates).
Finally, we note that
\begin{equation*} 
    \iF \iD\square_\eta\hat\beta 
    =2 \iF\delta_\eta\hat\beta
    =-2 \delta_\eta \iF\hat\beta 
    =0,
\end{equation*}
such that applying $m_f{}^l$ to $\square_\eta\hat\beta$ directly yields
\begin{align*}
    m_f{}^l(\square_\eta\hat\beta)
    &=\bigl[(j^{df}\iD + j^{dc}\iF - \Fd j^{dc}\iD + j^{dc} j^{df}\iF\iD)\square_\eta\hat\beta\bigr]_{|X_f}\\
    &=\bigl[(2j^{df}-j^{dc}(\LF + 2\Fd)) \delta_\eta\hat\beta\bigr]_{|X_f}.
\end{align*}
Then, noting that
    $(f\LF + 2\Fd)\delta_\eta\hat\beta
    = (f\LF - \Fd\LD)\delta_\eta\hat\beta
    = \L_{\sss{E_f}}\delta_\eta\hat\beta
    = \iEf d\delta_\eta\hat\beta$ and
$2j^{df}\delta_\eta\hat\beta 
    = -j^{df}\LF\delta_\eta\hat\beta
    = -j^{df}\iD d\delta_\eta\hat\beta$,
it reads also as
\begin{equation*}
    m_f{}^l(\square_\eta\hat\beta)
    =\bigl[-(j^{df}\iD+j^{dc}\iEf)d\delta_\eta\hat\beta]_{|X_f}.
\end{equation*}

Alternatively, we can write the longitudinal projection of $\square_\eta\hat\beta$ as
\begin{align*}
    m_f{}^l(\square_\eta\hat\beta)
    &=\bigl[(j^{df}\iD + j^{dc}\iF - \Fd j^{dc}\iD
        + j^{dc} j^{df}\iF\iD)\square_\eta\hat\beta\bigr]_{|X_f}\\
    &=\bigl[(-j^{df}\LD -j^{dc}\LF + \Fd j^{dc}\LD
        - j^{dc}j^{df}\iF\LD)\delta_\eta\hat\beta
        \bigr]_{|X_f}\\
    &= -m_f{}^l(d\delta_\eta\hat\beta),
\end{align*}
since, e.g., $\iD\square_\eta\hat\beta = -\LD\delta_\eta\hat\beta = -(\iD d + d\iD)\delta_\eta\hat\beta = -\iD d\delta_\eta\hat\beta$.

Finally, on $X_f$, since one has
    $m_f{}^*\square_\eta\hat\beta
    = \square_\eta\hat\beta 
    - m_f{}^l\square_\eta\hat\beta$,
this leads to
\begin{align*}
    \square_f\beta 
    &= \square_\eta\hat\beta
    + m_f{}^l(d\delta_\eta\hat\beta)\\
    &= \square_\eta\hat\beta
    + (j^{df}\iD + j^{dc}\iEf)d\delta_\eta\hat\beta,
\end{align*}
which proves the second formula [Eq.~\eqref{eq:square_fbeta}].
\end{proof}

\subsection{Proof of Prop.~\ref{prop:Riemann_f}}\label{proof:Riemann_f}

\begin{proof}
    This is a straightforward consequence of Gauss' equations, see Sec.~4.7 of Ref.~\onlinecite{jost2008riemannian} and Eq.~(4.7.7) in particular, in which the Riemann tensor is, in our case, given by
    \begin{equation}\label{eq:Riemann_ll}
        \Riemann_f 
        = \tfrac{1}{2}(\ell_{n+1}\owedge\ell_{n+1}-\ell_n\owedge\ell_n)
        = \tfrac{1}{2}(\ell_{n+1}-\ell_n)\owedge(\ell_{n+1} + \ell_n),
    \end{equation}
    with $\ell_n(u,v) = \ell_{e_n}(u,v) = \eta(\nabla^\eta_u e_n, v)$ and, similarly,
    $\ell_{n+1}(u,v) = \ell_{e_{n+1}}(u,v) = \eta(\nabla^\eta_u e_{n+1}, v)$ with 
    $e_n$ and $e_{n+1}$ the orthonormal vectors to $X_f$ in $\setR^{n+2}$, see Eqs.~\eqref{eq:en} and \eqref{eq:en+1}.
    First we note that
    \begin{align*}
        &\ell_{\sss D}(u,v)
        = \eta(\nabla^\eta_u D, v)
        = (u^\alpha\dr_\alpha y^\beta) v_\beta
        = u^\alpha \delta_\alpha{}^\beta v_\beta
        = \eta_{\alpha\beta} u^\alpha v^\beta
        = \eta(u,v),\\
        &\ell_{\sss F}(u,v)
        = \eta(\nabla^\eta_u F, v)
        = (u^\alpha\dr_\alpha (\dr_\beta f)) v^\beta
        = (\dr_\alpha\dr_\beta f) u^\alpha v^\beta
        = (\nabla^\eta df)(u,v).
    \end{align*}
    Then, from the definitions of $e_n$ and $e_{n+1}$ [Eqs.~\eqref{eq:en} and \eqref{eq:en+1}], we deduce that
    \begin{align*}
        &\ell_{n+1}-\ell_n = \ell_{\sss D} = \eta,\\
        &\ell_{n+1}+\ell_n = 2\ell_{\sss F} -\Fd\ell_{\sss D} 
            = 2(\nabla^\eta df) - \Fd\eta.
    \end{align*}
    Equation~\eqref{eq:Riemann_f} follows then from Eq.~\eqref{eq:Riemann_ll}.
    Equations~\eqref{eq:Ricci_f} and \eqref{eq:R_f} are then obtained from the contractions of the Riemann tensor, see Eqs.~\eqref{eq:C_13} and \eqref{eq:C_13g}.
\end{proof}

\subsection{Proofs of Lemma~\ref{lemma:1} and Prop.~\ref{prop: Weitzenbock}}\label{proof:Weitzenbock}

\begin{proof}(of Lemma~\ref{lemma:1}).
Following Eqs.~\eqref{eq: Weitzenbock} and \eqref{eq:Def_KN}  we have
\begin{equation*}
    \square-\Delta
     = (g_{ac}T_{bd}+g_{bd}T_{ac} 
        -g_{bc}T_{ad}-g_{ad} T_{bc})j^ai^bj^ci^d,
\end{equation*}
in which we compute each term.
First,
\begin{equation*}
  g_{ac}T_{bd} j^ai^bj^ci^d
    = g_{ac}T_{bd}(- j^aj^ci^bi^d+g^{bc}j^ai^d)
    =0+T_{ad}j^ai^d
    = {}^{\sss D} T,
\end{equation*}
and, similarly, $g_{bd}T_{ac}j^ai^bj^ci^d = {}^{\sss D} T$.
The third term gives
\begin{align*}
    -g_{bc}T_{ad}j^ai^bj^ci^d
    =-g_{bc}T_{ad} (-j^aj^ci^bi^d+g^{bc}j^ai^d)
    = (a-n-1)\, {}^{\sss{D}}T,
\end{align*}
with $a$ the degree of the differential form on which the operator is applied to,
and the fourth term gives
\begin{align*}
    -g_{ad}T_{bc}j^ai^bj^ci^d
    &=-g_{ad}T_{bc} (-j^aj^ci^bi^d+g^{bc}j^ai^d)\\
    &=g_{ad}T_{bc}j^aj^ci^bi^d-a\Tr(T)\\
    &=g_{ad}T_{bc}(-j^aj^ci^di^b)-a\Tr(T)\\
    &=g_{ad}T_{bc}(+j^ai^dj^ci^b-g^{cd}j^ai^b)-a\Tr(T)\\
    &=a T_{bc}j^ci^b-T_{ba}j^ai^b-a\Tr(T)\\
    &=(a-1){}^{\sss{D}}T -a\Tr(T),
\end{align*}
which gives the result.
\end{proof}

\begin{proof}(of Prop.~\ref{prop: Weitzenbock}).
 Equation~\eqref{eq:Riemmann_ff} together with lemma~\ref{lemma:1} gives
\begin{equation*}
    (\square_f-\Delta_f)\alpha
    =\Fd a(n-a)+(2a-n)\,{}^{\sss D}N_f-a\Tr(N_f),
\end{equation*}
using the fact that ${}^{\sss{D}}\eta_f\alpha = j^ai_a\alpha = a\alpha$.
In fact, we have $\Tr(N_f)=\square_\eta f$.
Indeed, on $X_f$ we have 
\begin{align*}
    \Tr(N_f)
    &= \eta^{\mu\nu}\nabla^\eta df(e_\mu,e_\nu)\\  
    &=\Tr(\nabla^\eta df)-\eta^{nn}\nabla^\eta df(e_n,e_n)-\eta^{(n+1)(n+1)}\nabla^\eta df(e_{n+1},e_{n+1})\\
    &=\square_\eta f-2\nabla^\eta df(D,F)+\Fd \nabla^\eta df(D,D).
\end{align*}
Moreover, a simple calculation in Cartesian coordinates in ambient space, using the homogeneity of $f$, shows that $(\nabla^\eta df)(D,F)=(\nabla^\eta df)(D,D)=0$.
This proves Eq.~\eqref{eq: LB1}.

In order to obtain Eq.~\eqref{eq: LB2}, we remark that $m_f{}^*\{(\nabla^\eta(df)_{\alpha\beta}j^\alpha i^\beta-\nabla^\eta(df)_{\mu\nu}j^\mu i^\nu)\alpha\}$   contains only terms with $j^n$ or $j^{n+1}$ on the left, which vanish thanks to $m_f{}^*$, or terms with $i^n$ or $i^{n+1}$ on the right also vanishing since $\alpha$ is transverse. Together with  Eq.~\eqref{eq:Sdphi} it gives Eq.~\eqref{eq: LB2}.
\end{proof}

\end{document}